\documentclass{aa}  
\usepackage{graphicx}
\usepackage{txfonts}
\usepackage{natbib}
\bibpunct{(}{)}{;}{a}{}{,}
\usepackage[figuresright]{rotating}
\usepackage{aalongtable,lscape}
\begin{document}

	\title{ORGANIC MOLECULES IN THE GALACTIC CENTER.\thanks{Table 7 is only
	available in electronic at the CDS via anonymous ftp to
	cdsarc.u-strasbg.fr (130.79.125.5) or via
	http://cdsweb.u-strasbg.fr/Abstract.html}}
   	\subtitle{HOT CORE CHEMISTRY WITHOUT HOT CORES.}

\author{M. A. Requena-Torres\inst{1}, J. Mart\'in-Pintado\inst{1}, 
	A. Rodr\'iguez-Franco,
          \inst{1}\inst{2}
	  	S. Mart\'in\inst{3}, N. J. Rodr\'iguez-Fern\'andez\inst{4}, 
		and \mbox{P. de Vicente}\inst{5}
		}
   \offprints{M. A. Requena-Torres, \email{requena@damir.iem.csic.es}}
   \institute{Departamento de Astrof\'isica Molecular e Infrarroja, Instituto de
   	Estructura de la Materia-CSIC
		C\ Serrano 121, E-28006 Madrid, Spain
		\email{requena@damir.iem.csic.es}
	\and
		Escuela Universitaria de \'Optica, Departamento de Matem\'atica
		Aplicada (Biomatem\'atica), Universidad Complutense de Madrid, 
		Avenida Arcos de Jal\'on s/n, E-28037 Madrid, Spain.
	\and
		Instituto de Radioastronom\'ia Milim\'etrica
		Av. Divina Pastora 7, Local 20, E-18012 Granada, Spain
		\email{martin@iram.es}
	\and
		Observatoire de Bordeaux, L3AB (UMR 5804), CNRS/Universit\'e
		Bordeaux 1, BP 89, 2 rue de l'Observatoire, 33270 Floirac,
		France \\
   		\email{nemesio.rodriguez@obs.u-bordeaux1.fr}
	\and
		Observatorio Astron\'omico Nacional,
		Centro Astron\'omico de Yebes
		Apartado 148, 19080 Guadalajara, Spain
		\email{p.devicente@oan.es}
	}

   \date{}
   
  \abstract
   {}
   {We study the origin of large abundances of complex organic molecules 
   in the Galactic center (GC).} 
  {We carried out a systematic study of the complex organic molecules 
CH$_3$OH, C$_2$H$_5$OH, (CH$_3$)$_2$O, HCOOCH$_3$, HCOOH, CH$_3$COOH, H$_2$CO,   
and CS toward 40 GC molecular clouds. Using the LTE approximation, we 
derived the physical properties of GC molecular clouds 
and the abundances of the complex molecules.}
  {The CH$_3$OH abundance between clouds varies by nearly two orders of 
magnitude from 2.4$\times$10$^{-8}$ to 1.1$\times$10$^{-6}$. 
The abundance of the other complex organic molecules relative to that of 
CH$_3$OH is basically 
independent of the CH$_3$OH abundance, with variations of only a factor 4--8. 
The abundances of complex organic molecules in the GC
are compared with those measured in hot cores and hot corinos, 
in which these complex molecules are also abundant. 
We find that both the abundance 
and the abundance ratios of the complex molecules
relative to CH$_3$OH in hot cores are similar to those found in the GC clouds.
However, hot corinos show different abundance ratios than observed in 
hot cores and in GC clouds. The rather 
constant abundance of all the complex molecules relative to CH$_3$OH 
suggests that all complex molecules are ejected from grain mantles by shocks. 
Frequent ($\sim$10$^5\,$years) shocks with velocities $>$6$\,$km~s$^{-1}$ are
required to explain the high abundances in gas phase of complex organic molecules in
the GC molecular clouds.
The rather uniform abundance ratios in the GC clouds and in Galactic hot cores
indicate a similar average composition of grain mantles in both kinds of regions.
The Sickle and the Thermal Radio Arches, affected by UV radiation, show different 
relative abundances in the complex organic molecules due to the differentially
photodissociation of these molecules.
}
   {}

\titlerunning{Organic molecules in the GC}
\authorrunning{Requena-Torres et al.}

   \maketitle
%
\section{Introduction}

Gas-grain interaction in the interstellar medium (ISM) can have a large impact 
on its chemistry because of the desorption and/or depletion of molecules onto
grains mantles. The gas phase abundance of molecules, which are believed to be 
efficiently formed on dust grains such as, methanol (CH$_3$OH), ethanol 
(C$_2$H$_5$OH), and formaldehyde (H$_2$CO), can be enhanced by orders of 
magnitude due to the ejection/evaporation of these molecules from grain
mantles.\\

One of the interesting features of the gas-phase chemistry after ejection of 
these molecules is the fast conversion of these parent molecules into daughter 
molecules. The relative abundance of dimethyl ether ((CH$_3$)$_2$O)
is expected to change by three orders of magnitude in a short period of time 
after CH$_3$OH is ejected to gas phase \citep{mil91,case93,char95,hor04}. 
Basically all models of alcohol-driven chemistry have indicated that 
(CH$_3$)$_2$O 
and methyl formate (HCOOCH$_3$) will reach their largest abundances 
$10^4$-$10^5$ years after the
ejection of CH$_3$OH into gas phase. However, \citet{hor04} have shown that  
HCOOCH$_3$ formation in gas phase from CH$_3$OH is less 
efficient than previously considered in the models. This casts some doubts on
the gas-phase production of HCOOCH$_3$, and they propose that this molecule
is also formed on grain mantles. The gas phase/grain formation of 
(CH$_3$)$_2$O has been studied by \citet{pee06}, showing that the main
path to form (CH$_3$)$_2$O is by gas-phase reactions and that its
formation on grains is a minor source of the observed abundances.
Other organic molecules like formic acid (HCOOH) can be produced in both gas 
phase and grain mantle chemistry \citep{liu02}. 

So far, large abundances of complex molecules have been detected in three
different kinds of objects:
hot cores associated with massive star formation  \citep{ike01}, hot
corinos associated with low-mass star formation 
\citep{bot06}, and the Galactic center (GC) clouds \citep{marpin01}. 
Their physical properties are very different. While the hot cores
and hot corinos are
small ($\lesssim$0.1$\,$pc), hot ($>$100$\,$K), and very 
dense ($\gtrsim$10$^{5}\,$cm$^{-3}$) condensations, the GC molecular clouds 
show averaged scales of 20-30~pc, kinetic temperatures $\sim$50-200~K, and 
an averaged density of $\sim$10$^4\,$cm$^{-3}$ \citep{gus04}.
The molecular gas in the inner region of our Galaxy presents a different
chemistry than in the Galactic disk \citep{marpin97}. In the GC, widespread large
abundances of grain-processed molecules like CH$_3$OH, C$_2$H$_5$OH, and 
silicon monoxide (SiO) are observed \citep{got79,min92,marpin97,marpin01,hut98}.
It has been proposed that the sputtering of grains and grain mantles produced by
widespread shocks with moderated velocities of $\le$40 km s$^{-1}$ are the
responsible of the 
$``$rich'' chemistry observed in the GC clouds. The origin of the large scale
shocks is so far unclear. These shocks could be produced by cloud-cloud
collisions associated with the large-scale dynamics in the context of a barred
potential \citep{has94,hut98,nem06}, by wind-blown bubbles 
driven by evolved massive stars \citep{marpin99}, or by
hydrodynamic (HD) or magneto hydrodynamic turbulence (MHD) \citep{mor96}. 
Since the formation of some complex organic molecules is
believed to proceed in gas phase after the passage of the shocks and their
abundances rapidly evolve, one could use the abundances of these molecules 
to gain insight into the chemistry of complex molecules in the ISM and to
constrain the age of shocks in the GC \citep{marpin01}.

Following these ideas we made a systematic study of 40 molecular clouds in 
the GC region, between Sgr~B2 and Sgr~C, in complex molecules believed to be
formed on grains like CH$_3$OH, C$_2$H$_5$OH, and HCOOCH$_3$, in gas phase from 
CH$_3$OH like (CH$_3$)$_2$O and in gas phase and/or grains like HCOOH and 
H$_2$CO.
We also searched for the less abundant isomer of HCOOCH$_3$, the acetic 
acid (CH$_3$COOH). As tracers of the total column density of the molecular gas,
we observed molecules like C$^{18}$O, $^{13}$CO, and CS. 
Our systematic study shows that all the complex organic molecules present 
similar relative abundances with respect to CH$_3$OH, except for the regions
where photodissociation could be important. Furthermore, the abundance and
abundance ratios of these complex molecules in the GC are similar to those 
observed in hot cores, while hot corino abundances seem to have similar 
ratios for different objects. 
Our results suggest that all 
complex organic molecules have been ejected from grain mantles and that their
abundances represent, in first approximation, the grain mantle composition.

\begin{table}
\begin{minipage}[t]{\columnwidth}
\caption{Molecular line parameters}             
\label{trans}      
\centering          
\renewcommand{\footnoterule}{}  
\begin{tabular}{l c r r r }     
\hline\hline       
Molecule & Transition & Frequency & $E_{\rm u}/\kappa$  & $\mu_x^2$S\\
&&(MHz)&(K)&\\
\hline                    
CH$_{3}$OH.......			&$3_{0}\rightarrow2_{0}\,$E	     	& 145093.75 &  27.06 & 2.38\\
					&$3_{-1}\rightarrow2_{-1}\,$E	      	& 145097.47 &  19.52 & 2.11\\
		  			&$3_{0}\rightarrow2_{0}\,$A+    	& 145103.23 &  13.94 & 2.38\\
					&$5_{0}\rightarrow4_{0}\,$E          	& 241700.22 &  47.95 & 3.91\\ 
					&$5_{-1}\rightarrow4_{-1}\,$E	      	& 241767.22 &  40.41 & 3.75\\ 
	      				&$5_{0}\rightarrow4_{0}\,$A+	     	& 241791.43 &  34.83 & 4.94\\ 
$^{13}$CH$_{3}$OH ...			&$3_{03}\rightarrow2_{02}\,$E		& 141595.48 &  26.71 & 2.43\\
 					&$3_{-13}\rightarrow2_{-12}\,$E	  	& 141597.06 &  19.21 & 2.16\\
					&$3_{03}\rightarrow2_{02}\,$A+		& 141602.53 &  13.60  & 2.43\\
C$_2$H$_5$OH ......                     &$4_{14}\rightarrow3_{03}$              & 90117.61  &  9.36  & 5.35\\
					&$7_{07}\rightarrow6_{16}$              & 104487.26 &  23.26 & 8.61\\
					&$9_{09}\rightarrow8_{18}$                      & 142285.05 &  37.17 & 12.66\\
HCOOCH$_3$ ..                           &$7_{25}\rightarrow6_{24}\,$E                    & 90145.69  &  19.69 & 17.00\\
					&$7_{25}\rightarrow6_{24}\,$A                    & 90156.48  &  19.67 & 17.00\\
					&$8_{08}\rightarrow7_{07}\,$E   	& 90227.61  &  20.09 & 20.99\\
					&$8_{08}\rightarrow7_{07}\,$A		   & 90229.63  &  20.07 & 20.99\\
					&$9_{45}\rightarrow8_{44}\,$E             	& 111408.48 &  37.27 & 19.13\\
					&$9_{18}\rightarrow8_{17}\,$E             	& 111674.10 &  28.15 & 23.12\\
					&$9_{18}\rightarrow8_{17}\,$A               	& 111682.19 &  28.13 & 23.12\\
(CH3)$_{2}$O\footnote{EE, AA, EA, and AE substates blended. Only the most intense transition is given.} ....      &$7_{26}\rightarrow7_{17}$            		    & 104703.30  &  31.07 & 5.43\\ 
					&$7_{07}\rightarrow6_{16}$                      & 111783.01 &  25.26 & 6.80\\ 
HCOOH .......                           &$4_{22}\rightarrow3_{21}$                      & 90164.25  &  23.53 & 5.78\\ 
					&$5_{05}\rightarrow4_{04}$                      & 111746.79 &  16.13 & 9.65\\ 
CH$_{3}$COOH ..                        &$8_{*8}\rightarrow7_{*7}\,$A                    & 90246.25  &  20.30  & 43.20\\ 
					&$8_{*8}\rightarrow7_{*7}\,$E                    & 90203.44  &  20.30  & 43.20\\ 
					&$10_{*10}\rightarrow9_{*9}\,$A                  & 111548.53 &  30.50  & 54.80\\ 
					&$10_{*10}\rightarrow9_{*9}\,$E                  & 111507.27 &  30.50  & 54.80\\ 
CS ...............                        &$3\rightarrow2$                                & 146969.03 &  14.12 & 11.56\\ 
					&$5\rightarrow4$                                & 244936.64 &  35.28 & 19.84\\ 
C$^{18}$O ........                      &$1\rightarrow0$                                & 109782.17 &  5.28 & 0.01\\ 
					&$2\rightarrow1$                                & 219560.35 &  15.82 & 0.02\\ 
$^{13}$CO .........                     &$1\rightarrow0$                                & 110201.35 &  5.30  & 0.01\\
H$_2^{13}$CO .......                      &$2_{02}\rightarrow1_{01}$                      & 141983.74 &   7.11 & 10.87\\
\hline                  
\end{tabular}
\end{minipage}
\end{table}

\begin{figure*}[]
\includegraphics[angle=270,width=17cm]{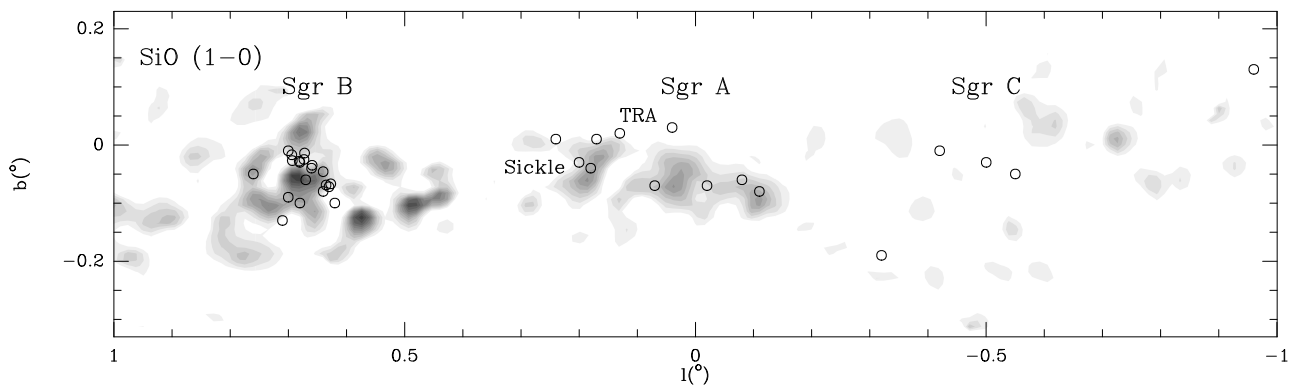}
\caption{Position of the observed sources superimposed on the SiO J=1$\rightarrow$0 maps
from the GC region by \citep{marpin97}. 
}\label{fig}
\end{figure*}
\section{Observations and results}
The observations of all transitions of the complex organic molecules shown in   
Table \ref{trans} were carried out with the IRAM 30-m radio telescope at Pico 
Veleta (Spain). The data were obtained in different seasons between 1996 and 
2003. The half-power beam width of the telescope was 24$''$, 17$''$, and 12$''$ 
for the 3, 2, and 1.3$\,$mm bands. The receivers, equipped with SIS mixers, were 
tuned to single sideband with image rejections $\ga\,$10$\,$dB. The typical 
system temperatures were 300, 500, and 900$\,$K for the 3, 2, and 1.3$\,$mm 
lines, respectively. We used two filterbanks of 256$\,$x$\,$1$\,$MHz and one of
512$\,$x$\,$1$\,$MHz as spectrometers. The velocity resolutions provided by the
filter banks were 3, 2, and 1.3 km s$^{-1}$ for the 3, 2, and 1.3$\,$mm bands,
respectively. Spectra were calibrated using the standard dual load system. 
We used the antenna temperature scale ($T_{\rm A}^*$) for the
line intensities because the emission is rather extended and completely fills
the beam.

Figure \ref{fig} shows the location of the sources we observed superimposed 
on the large-scale SiO map of the GC by \citet{marpin97}.  The 
sources were selected from the SiO maps of \citet{marpin97}.
Some of these sources had already been observed in C$_2$H$_5$OH by 
\citet{marpin01}. We also
included some clouds in Sgr~C, the Thermal Radio Arches (TRA) \citep{segu87},
and the Sickle \citep{segu91}, where 
the C$_2$H$_5$OH emission was not detected by \citet{marpin01}.  
The selected molecular clouds are spread over the region between Sgr~C and 
Sgr~B2. The positions of the sources are given in galactic and equatorial 
coordinates in Table \ref{sou}. The nomenclature used to designate them is 
similar to that of \citet{marpin01}: MC stands for Molecular 
Cloud followed by a G and the galactic coordinates, as recommended by the IAU.
We grouped the sources in different regions as shown in 
Fig. \ref{fig} and Table \ref{sou}.

\begin{table*}[]
\begin{minipage}[t]{180mm}
\caption{Source position}             
\label{sou}      
\centering          
\begin{tabular}{l c c c c l }     
\hline\hline       
source&l($^o$)&b($^o$)&$\alpha$ (B1950)&$\delta$ (B1950)&Region\\
\hline                    
MC G$-$0.96+0.13 	& $-$0.96 	& +0.13 	& 17$^{\rm h}$ 39$^{\rm m}$ 36$^{\rm s}$6    	& -29$^{\circ}$ 39$'$ 47$''$ & Sgr~C\\
MC G$-$0.55$-$0.05  	& $-$0.55 	& $-$0.05 	& 17$^{\rm h}$ 41$^{\rm m}$ 20$^{\rm s}$0 	& -29$^{\circ}$ 24$'$ 30$''$ & Sgr~C \\
MC G$-$0.50$-$0.03  	& $-$0.50 	& $-$0.03 	& 17$^{\rm h}$ 41$^{\rm m}$ 21$^{\rm s}$2 	& -29$^{\circ}$ 21$'$ 26$''$ & Sgr~C \\
MC G$-$0.42$-$0.01  	& $-$0.42 	& $-$0.01 	& 17$^{\rm h}$ 41$^{\rm m}$ 24$^{\rm s}$1 	& -29$^{\circ}$ 15$'$ 51$''$ & Sgr~C \\
MC G$-$0.32$-$0.19  	& $-$0.32 	& $-$0.19 	& 17$^{\rm h}$ 42$^{\rm m}$ 24$^{\rm s}$7 	& -29$^{\circ}$ 17$'$ 20$''$ & Sgr~C \\
MC G$-$0.11$-$0.08  	& $-$0.11 	& $-$0.08 	& 17$^{\rm h}$ 42$^{\rm m}$ 28$^{\rm s}$0 	& -29$^{\circ}$ 02$'$ 55$''$ & Sgr~A \\
MC G$-$0.08$-$0.06  	& $-$0.08 	& $-$0.06 	& 17$^{\rm h}$ 42$^{\rm m}$ 30$^{\rm s}$0 	& -29$^{\circ}$ 00$'$ 58$''$ & Sgr~A \\
MC G$-$0.02$-$0.07  	& $-$0.02	& $-$0.07 	& 17$^{\rm h}$ 42$^{\rm m}$ 40$^{\rm s}$0 	& -28$^{\circ}$ 58$'$ 00$''$ & Sgr~A \\
MC G+0.04+0.03  	& +0.04 	& +0.03 	& 17$^{\rm h}$ 42$^{\rm m}$ 26$^{\rm s}$2	& -28$^{\circ}$ 51$'$ 45$''$ & TRA \\
MC G+0.07$-$0.07  	& +0.07 	& $-$0.07 	& 17$^{\rm h}$ 42$^{\rm m}$ 54$^{\rm s}$2    	& -28$^{\circ}$ 53$'$ 30$''$ & TRA \\
MC G+0.13+0.02  	& +0.13 	& +0.02 	& 17$^{\rm h}$ 42$^{\rm m}$ 41$^{\rm s}$4    	& -28$^{\circ}$ 47$'$ 35$''$ & TRA \\
MC G+0.17+0.01  	& +0.17 	& +0.01 	& 17$^{\rm h}$ 42$^{\rm m}$ 50$^{\rm s}$0    	& -28$^{\circ}$ 45$'$ 50$''$ & TRA \\
MC G+0.18$-$0.04  	& +0.18 	& $-$0.04 	& 17$^{\rm h}$ 43$^{\rm m}$ 01$^{\rm s}$0    	& -28$^{\circ}$ 47$'$ 15$''$ & Sickle \\
MC G+0.20$-$0.03  	& +0.20 	& $-$0.03 	& 17$^{\rm h}$ 43$^{\rm m}$ 03$^{\rm s}$6    	& -28$^{\circ}$ 45$'$ 42$''$ & Sickle \\
MC G+0.24+0.01  	& +0.24 	& +0.01 	& 17$^{\rm h}$ 42$^{\rm m}$ 59$^{\rm s}$6    	& -28$^{\circ}$ 42$'$ 35$''$ & Sickle \\
MC G+0.62$-$0.10  	& +0.62 	& $-$0.10 	& 17$^{\rm h}$ 44$^{\rm m}$ 18$^{\rm s}$0    	& -28$^{\circ}$ 26$'$ 30$''$ & Sgr~B2\_a \\
MC G+0.64$-$0.08  	& +0.64 	& $-$0.08 	& 17$^{\rm h}$ 44$^{\rm m}$ 17$^{\rm s}$5    	& -28$^{\circ}$ 24$'$ 30$''$ & Sgr~B2\_a \\
MC G+0.67$-$0.06  	& +0.67 	& $-$0.06 	& 17$^{\rm h}$ 44$^{\rm m}$ 18$^{\rm s}$0    	& -28$^{\circ}$ 22$'$ 30$''$ & Sgr~B2\_a \\
MC G+0.68$-$0.10  	& +0.68 	& $-$0.10 	& 17$^{\rm h}$ 44$^{\rm m}$ 27$^{\rm s}$2    	& -28$^{\circ}$ 23$'$ 20$''$ & Sgr~B2\_a \\
MC G+0.70$-$0.01  	& +0.70 	& $-$0.01 	& 17$^{\rm h}$ 44$^{\rm m}$ 10$^{\rm s}$0    	& -28$^{\circ}$ 19$'$ 30$''$ & Sgr~B2\_a \\
MC G+0.70$-$0.09  	& +0.70 	& $-$0.09 	& 17$^{\rm h}$ 44$^{\rm m}$ 27$^{\rm s}$2    	& -28$^{\circ}$ 22$'$ 05$''$ & Sgr~B2\_a \\
MC G+0.71$-$0.13  	& +0.71 	& $-$0.13 	& 17$^{\rm h}$ 44$^{\rm m}$ 38$^{\rm s}$4    	& -28$^{\circ}$ 22$'$ 25$''$ & Sgr~B2\_a \\
MC G+0.76$-$0.05  	& +0.76 	& $-$0.05 	& 17$^{\rm h}$ 44$^{\rm m}$ 27$^{\rm s}$2    	& -28$^{\circ}$ 17$'$ 35$''$ & Sgr~B2\_a \\
SGR~B2N       		& +0.68 	& $-$0.03 	& 17$^{\rm h}$ 44$^{\rm m}$ 10$^{\rm s}$6    	& -28$^{\circ}$ 21$'$ 17$''$ & Hot Core \\
SGR~B2M       		& +0.66 	& $-$0.04 	& 17$^{\rm h}$ 44$^{\rm m}$ 10$^{\rm s}$6    	& -28$^{\circ}$ 22$'$ 05$''$ & Hot Core \\
MC G+0.694$-$0.017	& +0.694	& $-$0.017	& 17$^{\rm h}$ 44$^{\rm m}$ 10$^{\rm s}$0    	& -28$^{\circ}$ 20$'$ 05$''$ & Sgr~B2 \\
MC G+0.693$-$0.027	& +0.693	& $-$0.027	& 17$^{\rm h}$ 44$^{\rm m}$ 12$^{\rm s}$1    	& -28$^{\circ}$ 20$'$ 25$''$ & Sgr~B2 \\
MC G+0.627$-$0.067	& +0.627	& $-$0.067	& 17$^{\rm h}$ 44$^{\rm m}$ 12$^{\rm s}$1    	& -28$^{\circ}$ 25$'$ 05$''$ & Sgr~B2 \\
MC G+0.630$-$0.072	& +0.630	& $-$0.072	& 17$^{\rm h}$ 44$^{\rm m}$ 13$^{\rm s}$6    	& -28$^{\circ}$ 25$'$ 05$''$ & Sgr~B2 \\
MC G+0.672$-$0.014	& +0.672	& $-$0.014	& 17$^{\rm h}$ 44$^{\rm m}$ 06$^{\rm s}$1    	& -28$^{\circ}$ 21$'$ 05$''$ & Sgr~B2 \\
MC G+0.640$-$0.046	& +0.640	& $-$0.046	& 17$^{\rm h}$ 44$^{\rm m}$ 09$^{\rm s}$1    	& -28$^{\circ}$ 23$'$ 45$''$ & Sgr~B2 \\
MC G+0.635$-$0.069	& +0.635	& $-$0.069	& 17$^{\rm h}$ 44$^{\rm m}$ 13$^{\rm s}$6    	& -28$^{\circ}$ 24$'$ 45$''$ & Sgr~B2 \\
MC G+0.659$-$0.035	& +0.659	& $-$0.035	& 17$^{\rm h}$ 44$^{\rm m}$ 09$^{\rm s}$1    	& -28$^{\circ}$ 22$'$ 25$''$ & Sgr~B2 \\
MC G+0.681$-$0.028	& +0.681	& $-$0.028	& 17$^{\rm h}$ 44$^{\rm m}$ 10$^{\rm s}$6    	& -28$^{\circ}$ 21$'$ 05$''$ & Sgr~B2 \\
MC G+0.673$-$0.025	& +0.673	& $-$0.025	& 17$^{\rm h}$ 44$^{\rm m}$ 09$^{\rm s}$1    	& -28$^{\circ}$ 21$'$ 25$''$ & Sgr~B2\\ 
\hline                
\end{tabular}
\end{minipage}
{\footnotesize{Note. Galactic and equatorial coordinates of the selected 
sources. 
Nomenclature, as recommended by the IAU: MC followed by a G and the galactic 
coordinates, where MC stands for 
Galactic center molecular cloud.} We have grouped the sources by regions: sources
around Sgr~C, Sgr~A, the TRA, the Sickle, Sgr~B2 (\_a is to difference two  
observation sessions), and the  hot cores. Not all sources in the TRA or in 
the Sickle are affected by the same physical conditions.}
\end{table*}

\begin{figure*}[]
\includegraphics[angle=270,width=18cm]{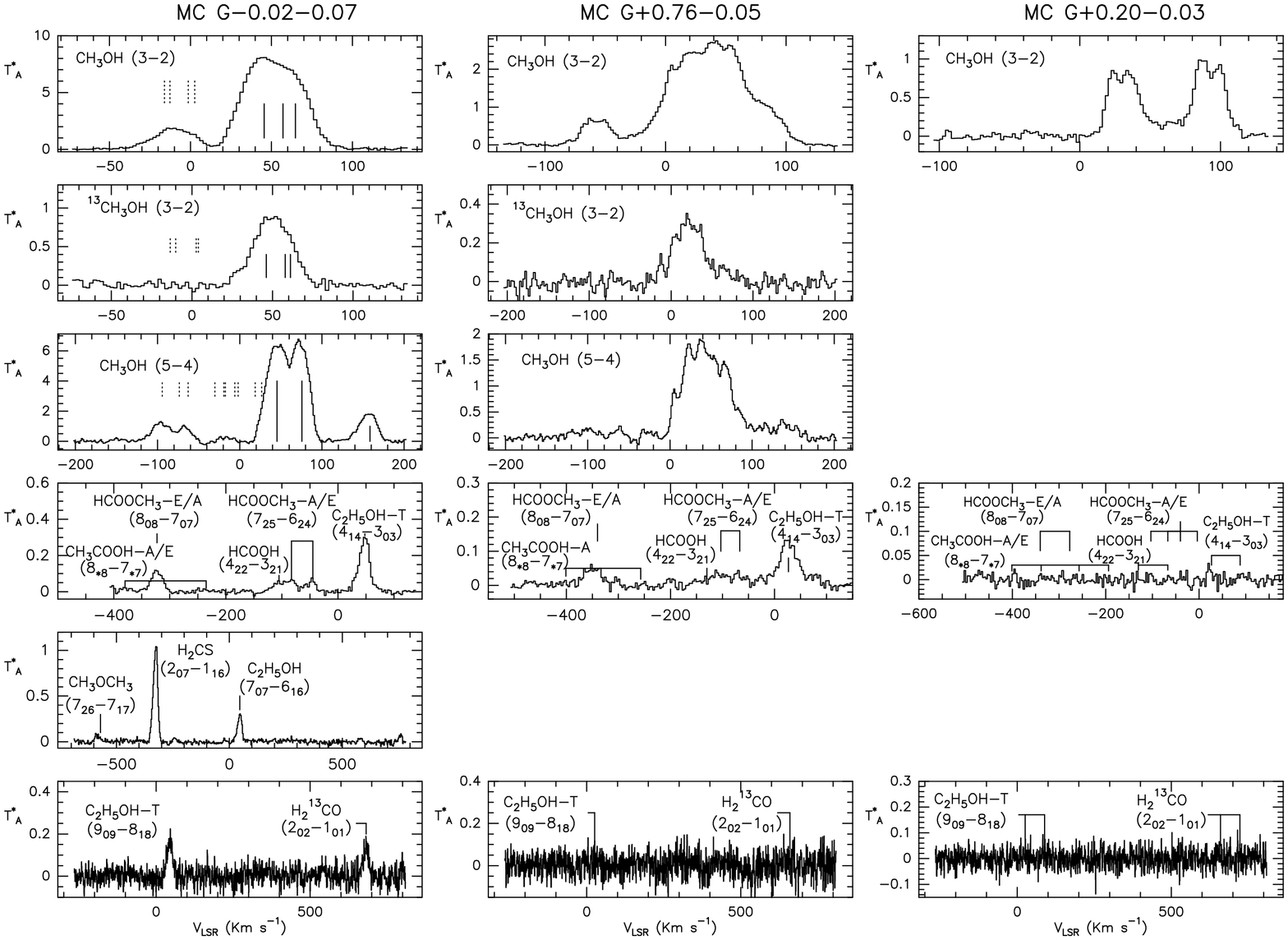}
\caption{Observed line profiles toward the sources \mbox{MC G$-$0.02$-$0.07},
\mbox{MC G+0.76$-$0.05}, and \mbox{MC G+0.20$-$0.03}. Line intensities are
expressed in antenna temperature (K), and velocities refer to LSR 
(km s$^{-1}$). The CH$_3$OH profiles are composed of several   
overlapped transitions arising from levels with different energies; in the
\mbox{MC G$-$0.02$-$0.07} source, we show the transitions with highest
intensities as vertical solid lines and the transitions with lowest intensities 
as dotted lines. These transitions are:
for CH$_3$OH J=$3$$\rightarrow$$2$, from left to right, 
$3_{2}$$\rightarrow$$2_{2}$A, 
$3_{1}$$\rightarrow$$2_{1}$E, $3_{-2}$$\rightarrow$$2_{-2}$E,
$3_{2}$$\rightarrow$$2_{2}$E, $3_{2}$$\rightarrow$$2_{2}$A-, 
$3_{0}$$\rightarrow$$2_{0}$A+, $3_{-1}$$\rightarrow$$2_{-1}$E, and 
$3_{0}$$\rightarrow$$2_{0}$E.
For the  $^{13}$CH$_3$OH J=$3$$\rightarrow$$2$, from left to right, 
$3_{21}$$\rightarrow$$2_{20}$A+, $3_{12}$$\rightarrow$$2_{11}$E,
$3_{21}$$\rightarrow$$2_{21}$E, $3_{-22}$$\rightarrow$$2_{-21}$E, 
$3_{22}$$\rightarrow$$2_{21}$A-, $3_{03}$$\rightarrow$$2_{02}$A+, 
$3_{-13}$$\rightarrow$$2_{-12}$E, and $3_{03}$$\rightarrow$$2_{02}$E.
For the CH$_3$OH J=$5$$\rightarrow$$4$ from left to right; 
$5_{2}$$\rightarrow$$4_{2}$E, $5_{-2}$$\rightarrow$$4_{-2}$E,
$5_{2}$$\rightarrow$$4_{2}$A+, $5_{1}$$\rightarrow$$4_{1}$E, 
$5_{-3}$$\rightarrow$$4_{-3}$E, $5_{3}$$\rightarrow$$4_{3}$E, 
$5_{2}$$\rightarrow$$4_{2}$A-, $5_{3}$$\rightarrow$$4_{3}$A-,
$5_{3}$$\rightarrow$$4_{3}$A+, $5_{4}$$\rightarrow$$4_{4}$E, 
$5_{-4}$$\rightarrow$$4_{-4}$E, $5_{4}$$\rightarrow$$4_{4}$A-, 
$5_{4}$$\rightarrow$$4_{4}$A, $5_{0}$$\rightarrow$$4_{0}$E, 
$5_{-1}$$\rightarrow$$4_{-1}$E, and  $5_{0}$$\rightarrow$$4_{0}$E. 
All these transitions
have been used for the multi-Gaussian fit. The spectrum toward 
\mbox{MC G+0.20$-$0.03} shows two velocity components.}\label{spe1}
\end{figure*}
\begin{figure*}[]
\includegraphics[angle=270,width=18cm]{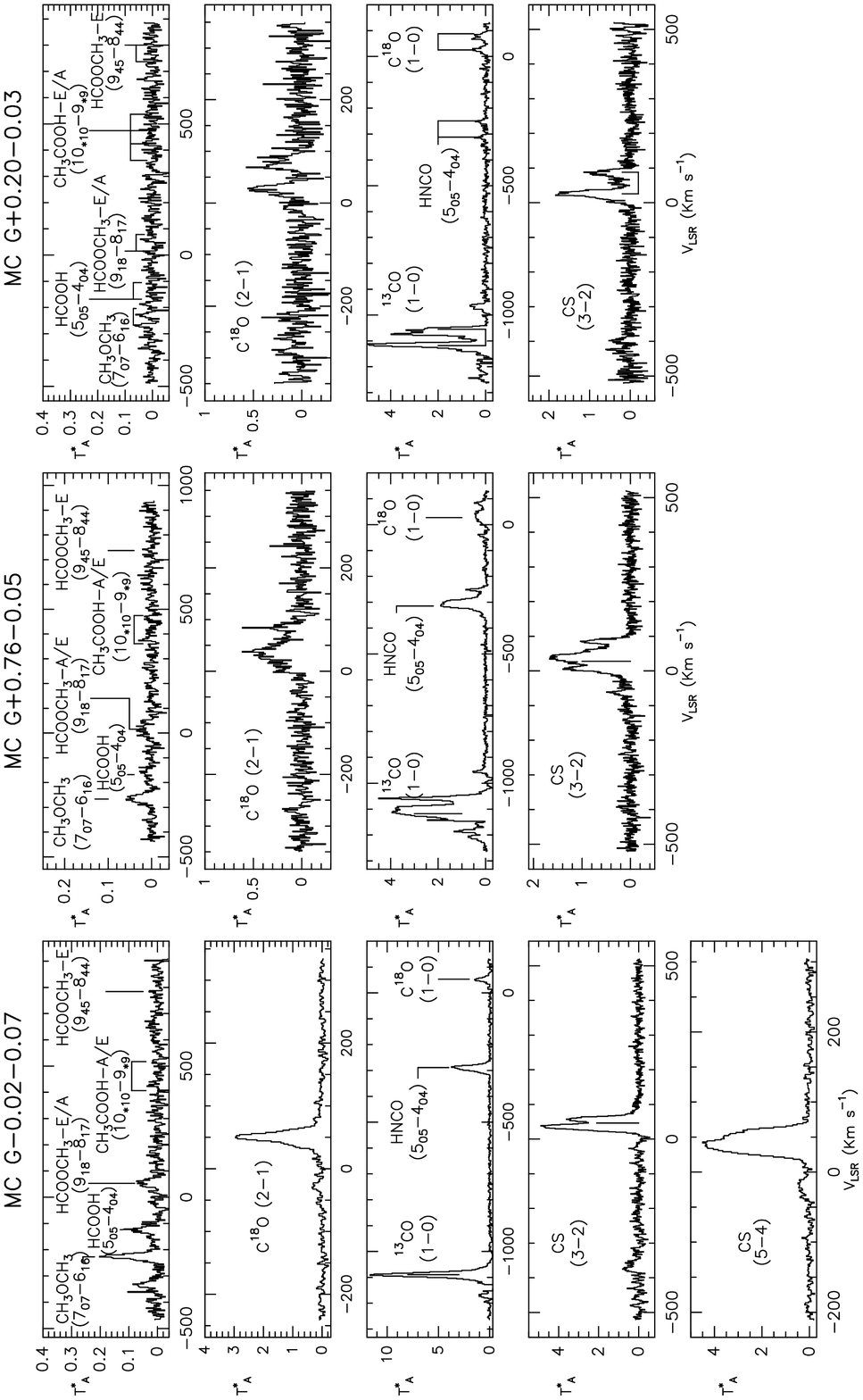}
\caption{Same as Fig. \ref{spe1}. The $^{13}$CO, C$^{18}$O, and CS profiles 
show wider lines than those of the complex organic molecules in Fig.
\ref{spe1}.}\label{spe2} 
\end{figure*}

Figures \ref{spe1} and \ref{spe2} show samples of line profiles for all 
observed molecules toward selected sources. Only for some sources like
\mbox{MC G$-$0.02$-$0.07} we did observe all the molecular lines.  The line
profiles of molecules such as CH$_3$OH are different from those of the CO
isotopomers and CS
due to the blending of several transitions. The complex organic molecules were 
only detected in some of the velocity components observed in C$^{18}$O, 
$^{13}$CO, and CS. A good example of this difference in line profiles is shown
in Figs. \ref{spe1} and \ref{spe2} for the source \mbox{MC G+0.76$-$0.05} where 
C$^{18}$O, $^{13}$CO, and CS show three or more velocity components, but
C$_2$H$_5$OH only shows one. Each velocity component has been treated as an 
independent molecular cloud. Multiple velocity components are also found toward 
\mbox{MC G+0.13+0.02}, \mbox{MC G+0.18$-$0.04}, and \mbox{MC G$-$0.32$-$0.19}.

\section{Analysis of the data}

To derive the physical and chemical properties of the selected molecular clouds,
we fitted Gaussian profiles 
to all detected molecular lines for each velocity component. The observed 
spectra of some organic molecules are more complex than a single Gaussian
profile because of the overlap of several transitions due to the large
linewidths observed 
toward the GC clouds. To account for the overlap, we used the following
constrains for the Gaussian fitting depending on the observed transitions:  

-- For molecules without internal rotation, like CS, C$^{18}$O, $^{13}$CO,
HCOOH, and H$_2^{13}$CO, and for molecules with internal rotation when the 
transitions are sufficiently separated, like those of C$_2$H$_5$OH and some of
the HCOOCH$_3$ lines, simple Gaussian profiles were fitted to each 
transition for every velocity component. 
As mentioned before, the CS, C$^{18}$O, and $^{13}$CO profiles generally show 
more velocity components than those observed in the complex organic molecules. 
For this work we only fitted the components in the velocity range where 
at least one organic molecule was detected. In the case where the velocity 
components overlapped we calculated the integrated intensity of CS, 
C$^{18}$O, and $^{13}$CO in the velocity range where the complex organic 
molecules show the emission. \\ 
\begin{sidewaystable*}
\begin{minipage}[t]{\textwidth}
\caption{Dipole moments and rotational constants of the observed
 molecules}             
\label{par}      
\centering          
\begin{tabular}{l c c c c c c c l l }     
\hline\hline       
molecule & $\mu_a$ & $\mu_b$ &
 A & B & C& Q& comments& references\\
  & (Debye)  & (Debye)  &  (MHz)  & (MHz)  & (MHz) &  & & \\ 
\hline                  
CH$_{3}$OH        &0.885&1.440  &127484.00&24679.98&23769.70&$1.28~T_{\rm rot}^{1.5}$    &Rotational levels split into E and A substates.				&\citet{men86}\\
		  &     &       &         &        &         &                      &Observed transitions with several blended lines.  		     		&\citet{and90a}\\
$^{13}$CH$_{3}$OH &0.899&1.440  &127527.40&24083.50&23197.00&$1.28~T_{\rm rot}^{1.5}$    &Rotational levels split into E and A substates.				&\citet{and90b}\\
		  &     &       &         &        &         &                      &Observed transitions with several blended lines.  		     		&\citet{xul97}\\
C$_2$H$_5$OH      &0.046&1.438  & 34891.77& 9350.68&  8135.34&$3.2~T_{\rm rot}^{1.5}$    &Rotational levels split into Trans and Gauche substates. 		&\citet{pea95,pea97}\\
		  &     &       &         &        &         &                      &Only Trans Transitions are exited at $\sim\,10\,K$. 	     		&\\
HCOOCH$_3$        &1.630&0.680  & 19985.71& 6914.75& 5304.48&$12.45~T_{\rm rot}^{1.5}$   &Rotational levels split into E and A substates.				&\citet{plu84,plu86}\\
		  &     &       &         &        &         &                      &  								     		&\citet{kua96}\\
(CH$_3$)$_{2}$O   &0.000&1.302  & 38788.20&10056.50& 8886.80&$91.6~T_{\rm rot}^{1.5}$    &Rotational levels spilt into AA, EE, AE and EA substates. 	       		&\citet{gro98}\\ 
		  &     &       &         &        &         &                      &$g_i$: (K$_a$K$_c$: $ee \leftrightarrow oo$) AA:EE:AE:EA=6:16:2:4 		&\\
		  &     &       &         &        &         &                      &$g_i$: (K$_a$K$_c$: $oe \leftrightarrow eo$) AA:EE:AE:EA=10:16:6:4		&\\
HCOOH             &1.396&0.260  & 77512.25&12055.11&10416.12&$1.68~T_{\rm rot}^{1.5}$    &                                                                       	&\citet{wil80}\\
		  &     &       &         &        &         &                       &                                                      	                 	& \citet{liu02}\\ 
CH$_{3}$COOH      &0.860&1.470  & 11335.58& 9478.73& 5324.99&$14.1~T_{\rm rot}^{1.5}$    &Rotational levels split into E and A substates.                          	&\citet{wlo88}\\
 		  &     &       &         &        &         &                      &  								     		&\citet{meh97}\\
H$^{13}_2$CO      &2.332&&281993.04&37809.11&33215.94&$0.57~T_{\rm rot}^{1.5}$   & Para (K$_a$=e) and Orto (K$_a$=o) split, $g_i\rightarrow$ 1(p)-3(o).	&\citet{joh72}\\
CS                &1.957&&  &24495.56& &$0.84~T_{\rm rot}$          &                                                                       	&JPL\\ 
C$^{18}$O         &0.111&&  &54891.42& &$0.38~T_{\rm rot}$          &                                                                       	&JPL\\ 
$^{13}$CO         &0.111&&  &55101.01& &$0.38~T_{\rm rot}$          &                                                                       	&JPL\\ 
\hline                  
\end{tabular}
\vfill
\end{minipage}
\end{sidewaystable*}

-- The CH$_3$OH and $^{13}$CH$_3$OH profiles are composed of several overlapped 
transitions arising from levels at different energies. In this case, we
fitted the blended transitions with multi-Gaussian profiles forced to have
the same linewidth and velocity separations corresponding to the rest
frequencies of the transitions. We also forced the relative intensities 
derived from the 
spectroscopic parameters (Tables \ref{trans} and \ref{par}), assuming the same rotational 
temperature for all the transitions. In addition to the radial velocity and 
linewidth, we also fitted the rotational temperature and the optical depths 
that match the observed profiles. In this case, the antenna temperatures were
obtained from ($T_{\rm rot}$-$T_{\rm bg}$)(1-$e^{-\tau}$). 
We used the optically thin emission from $^{13}$CH$_3$OH transitions to 
test the optical depths and the antenna temperatures obtained from the CH$_3$OH
fits. The optical depths derived from the CH$_3$OH 
lines are in good agreement with those obtained from the
[CH$_3$OH/$^{13}$CH$_3$OH] line-intensity ratios, assuming  a $^{12}$C/$^{13}$C
ratio of 20 \citep{wil94}. For the $^{13}$CH$_3$OH, and in
some cases for the CH$_3$OH lines, only the transitions with the largest
intensities were fitted.
In the case of the CH$_3$OH 5$_{0}$$\rightarrow$4$_{0}E$ 
transition, a simple Gaussian fit was made since this line is not 
overlapped with any other transition.\\
 
-- Due to its internal rotation, the (CH$_3$)$_2$O rotational levels are split 
into four substates with similar Einstein coefficients but different intensities
due to the nuclear spin degeneracy. These substates are called AA, EE, EA, and AE. Since 
the four substates are blended, we measured the integrated intensity of the
four substates and then summed all the spin weights of each substate to derive
the (CH$_3$)$_2$O column densities. \\

-- To obtain the upper limits to the column densities of undetected transitions
like those of CH$_3$COOH, we used the $3\sigma$ level for the integrated 
intensities.\\

Derived parameters from the Gaussian fits (peak intensity, integrated
intensity, radial velocity and linewidth) for all transitions in each source
are shown electronically in Table \ref{obs}, available at the CDS. 

The linewidths of the complex organic molecules in our sources are  
$\sim$15$\,$km s$^{-1}$, typical in the GC region. 
Radial velocities of the different sources range from $-$93$\,$km s$^{-1}$ to
140$\,$km s$^{-1}$.

\section{Derived parameters}
\subsection{Column densities, excitation temperatures, and densities}

Molecular column densities were derived by assuming optically thin
emission and the local thermodynamic equilibrium (LTE) approximation. In the
case of optically thick lines like those of CH$_3$OH, the optically thin
isotopic substitution was used. Under these conditions, the total column 
density, N, of a molecule is given by:

\begin{equation}
N=\frac{N_{\rm u}}{g_{\rm u}g_{\rm i}}Q e^{(E_{\rm u}/\kappa T_{\rm rot})} \label{dtot}
\end{equation}
 \begin{equation}
N_{\rm u}=1.67 \times10^{14}\frac{g_{\rm u}}
{\nu \mu^{2}S}\int{T^{*}_{\rm A}d{\rm v}}\,, \label{dsup}
\end{equation}
where N$_u$ is the column density in the upper level of the observed transition
in cm$^{-2}$,
$g_u$ the upper level degeneracy (2J$_{\rm u}$+1), $\nu$ the line frequency in  
GHz, $\mu$ the dipole moment in Debye, S the line strength, $\int{T^{*}_{\rm
A}d{\rm v}}$ the integrated intensity in K km s$^{-1}$, $g_i$ the spin weight
degeneracy, Q the partition function, $E_{\rm u}/\kappa$ the upper level energy
in K, and $T_{\rm rot}$ the rotational temperature also in K. Table \ref{par} 
summarizes the 
dipole moments, the rotational constants, the partition functions, and the 
statistical weights used to calculate the total column densities.

\subsubsection{Rotational temperatures}
\begin{figure}[h!]
\includegraphics[width=8cm]{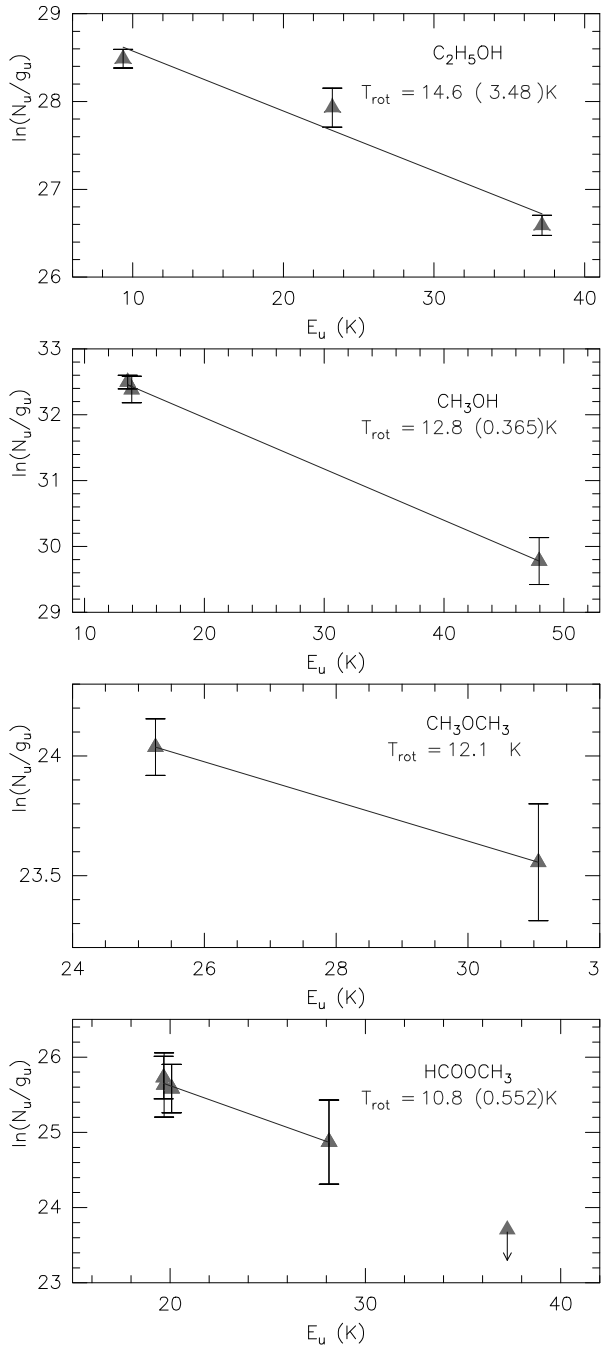}
\caption{Population diagrams for some complex molecules toward \mbox{MC
G$-$0.02$-$0.07}. The error bars are derived by assuming a calibration
uncertainty of 10\% and the Gaussian fit error.
Parentheses () indicate the calculated errors. Triangles with arrows represent 
upper limits to HCOOCH$_3$ column densities.}\label{temp}
\end{figure}

To derive the total column densities one first needs to estimate $T_{\rm rot}$. 
For 
molecules with more than one observed transition, we calculated $T_{\rm rot}$ 
from population diagrams. \mbox{Figure \ref{temp}} shows, as an example, the population 
diagrams ($ln(N_{\rm u}/g_{\rm u})$ vs. $E_{\rm u}/\kappa$) of some of the 
organic complex molecules derived for the source 
\mbox{MC G$-$0.02$-$0.07}.
In general, $T_{\rm rot}$ from different molecules (third column on Table
\ref{sou2}) agree within a factor
of 2. We find $T_{\rm rot}$ values between 6$\,$K and 18$\,$K, much lower 
than the typical kinetic temperatures of $\ga$100$\,$K  derived from NH$_3$ by 
\citet{hut93a} and H$_2$ by \citet{nem00,nem01a},
indicating subthermal excitation. We do not find any systematic trend in
$T_{\rm rot}$  as a function of the molecule. For sources with only one observed 
transition, we assumed an average value of $T_{\rm rot}\sim$8$\,$K. 

\subsubsection{Densities}
Once the $T_{\rm rot}$ is derived, we can make a rough estimate of the H$_2$ 
densities, providing that the kinetic temperatures and the collisional cross
section are known. The H$_2$ densities can be estimated 
from 
\begin{equation}
n=\frac{A_{\rm ul}}{<\rm v\sigma>_{\rm ul}}\frac{1+\left[1-exp{\left(\frac{h\nu}{\kappa
T_{\rm rot}}\right)}\right]\frac{1}{exp{\left(\frac{h\nu}
{\kappa T_{\rm bg}}\right)}-1}}
{exp{\left(\frac{h\nu}{\kappa T_{\rm rot}}\right)}exp{\left(-\frac{h\nu}{\kappa
T_{\rm k}}\right)}-1} \label{densi}
\end{equation}
where $A_{ul}$ is the Einstein coefficient between the upper and the lower
levels, $<$$\rm v\sigma$$>_{\rm ul}$ the collisional rate coefficient between the same
levels, $\nu$ the transition frequency, $h$ the Plank constant, $\kappa$ the 
Boltzmann constant, $T_{\rm rot}$ the rotational temperature, $T_{\rm bg}$ the 
background temperature, and $T_{\rm k}$ the kinetic temperature.
\citet{pot04} have calculated the collision rate coefficients 
($<$$\rm v\sigma$$>$) between E-CH$_3$OH and para-H$_2$ for kinetic temperatures 
between 5$\,$K and 200$\,$K. Assuming a kinetic temperature of 100$\,$K 
\citep{hut93a,nem01a}, we obtain H$_2$ densities of 
$\sim$10$^5\,$cm$^{-3}$ for the rotational
temperature derived from CH$_3$OH of 6--18$\,$K. Similar H$_2$ densities are also derived when the 
measured line intensities of E-CH$_3$OH are
fitted using the large velocity gradient (LVG) approximation model, provided by
J. Cernicharo, for the excitation of E-CH$_3$OH.  
We also estimated the H$_2$ densities from the excitation temperatures 
derived from C$_2$H$_5$OH. The collisional cross section in this case 
was estimated from
those of CH$_3$OH but corrected for the difference in mass and size. These
roughly estimated H$_2$ densities are also $\sim$10$^5\,$cm$^{-3}$.

The derived H$_2$ densities are based on the collisional rates for a kinetic 
temperature of 100~K. However, the linewidths in the GC molecular clouds
indicate the presence of supersonic turbulence and thus of non-equilibrium processes. 
In this case the collisional rates will not be described
by the kinetic temperatures. However, for the observed transitions the rate 
coefficients do depend weakly on the increasing collisional velocities 
\citep{pot04}, therefore our estimated densities will only be marginally
affected by the effect of a
larger velocity difference between the collisional partners than those 
represented by the kinetic temperature.

\subsubsection{The {\rm H$_2$} column densities}
To estimate the H$_2$ column density, we used the J=1$\rightarrow$0 and 
J=2$\rightarrow$1 C$^{18}$O and the J=1$\rightarrow$0 $^{13}$CO transitions 
using the LTE approximation. We used the isotopic ratios of 
$^{12}$C/$^{13}$C$=20$, $^{16}$O/$^{18}$O$=250$ \citep{wil94}, and a CO 
relative abundance of 
CO/H$_2=10^{-4}$ \citep{fre82}. The last column in Table \ref{sou2} 
shows the derived 
H$_2$ column densities. Our H$_2$ column densities, derived from the LTE 
approximation, are within a factor of $\le$2 of those derived by 
\citet{nem01a} using the LVG approximation. 

Table \ref{sou2} also shows the relative CS abundance estimated from the CS 
column densities derived using the $T_{\rm rot}$ obtained from CH$_3$OH.
Since our data refer to the line emission of the main CS isotopomer, optical 
depth effects might affect the derived CS column densities 
\citep{hut93b}. We compared our results with those derived 
from $^{13}$CS by \citet{mar06} in several common sources. The CS column
densities derived from $^{13}$CS are always higher than those of the CS by a
factor of 1-5. When the $^{13}$CS determinations are available, Table \ref{sou2}
shows the two values estimated for the relative abundance of CS. 

\subsubsection{Column densities  of complex organic molecules}
To derive the column densities of the complex organic molecules we used 
the parameters in Table \ref{par} and Eqs. \ref{dtot} and \ref{dsup}. 
The relative 
abundance of complex molecules derived from their column densities and the 
H$_2$ column densities are shown in Table \ref{sou2}. 

\subsection{Abundances and abundance ratios}
Table \ref{sou2} shows the derived rotation temperatures, fractional abundances,
and the H$_2$ column densities for the different molecular clouds.
Different sources in the same line of sight are identified 
by their radial velocities.
The fractional abundances of the complex organic molecules are very high, even
higher than in the hot cores. 
The abundances we find reach values up to 1.1$\times$$10^{-6}$ for 
CH$_3$OH, 6$\times$$10^{-8}$ for C$_2$H$_5$OH, 5.6$\times$$10^{-8}$ for 
(CH$_3$)$_2$O, 7.5$\times$$10^{-8}$ for HCOOCH$_3$, 4.1$\times$$10^{-9}$ for 
HCOOH, and 1.9$\times$$10^{-8}$ for H$_2$CO. 
We have only obtained upper limits to the abundance of CH$_3$COOH between 
$\le$5$\times$$10^{-10}$ and $\le$5.5$\times$$10^{-8}$.

There are some sources for which only upper limits to the abundances were 
measured for all the complex organic molecules except for CH$_3$OH. In some
cases, like the sources with galactic longitude  $\le$$-$0.30$^\circ$, the upper
limits to the abundances of all complex molecules are not relevant because of 
the lack of sensitivity of our observations. In this case the upper limits have
not been included in the discussion. Other cases, like the C$_2$H$_5$OH
upper limits toward \mbox{MC G+0.18$-$0.04}, \mbox{MC G+0.20$-$0.03} (Sickle), and 
\mbox{MC G+0.13+0.02} (TRA) are relevant and therefore included in our
discussion.
The molecular emission toward the Sickle shows 
two velocity components at $\sim$25 km s$^{-1}$ and $\sim$80 km s$^{-1}$, 
which are believed to be affected by the UV radiation produced by the Quintuplet
cluster \citep{nem01b}.
The molecular emission toward \mbox{MC G+0.13+0.02}, located in the TRA, also
seems to be affected by the UV radiation from the
Arches Cluster. In fact only the -30 km s$^{-1}$ velocity component, which is 
clearly associated with the ionized gas \citep{segu87}, shows lower 
C$_2$H$_5$OH abundance than in other sources.

\begin{figure}[h!]
\includegraphics[width=8.05cm]{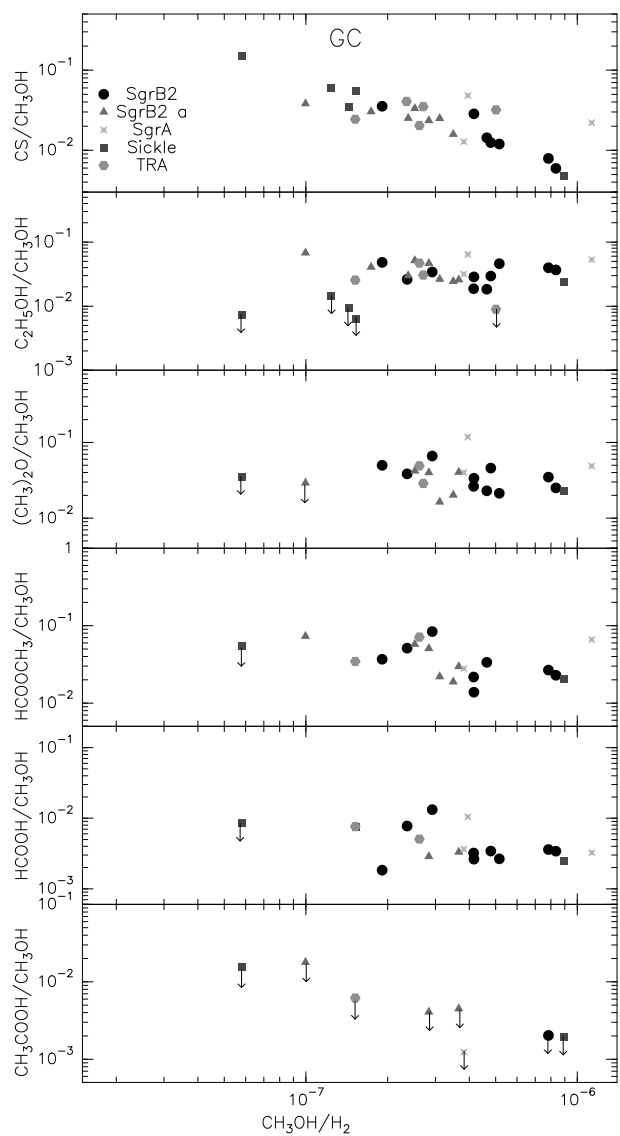}
\caption{Relative abundances of organic complex molecules with respect to 
CH$_3$OH as a function of the CH$_3$OH abundance for the selected GC clouds. 
Different
regions are represented with different symbols (see Table \ref{sou}). The 
arrows represent upper limits. Only the detections and significant 
upper limits are shown.\label{abun1}}
\end{figure}
\begin{figure}[h!]
\includegraphics[angle=270,width=8.05cm]{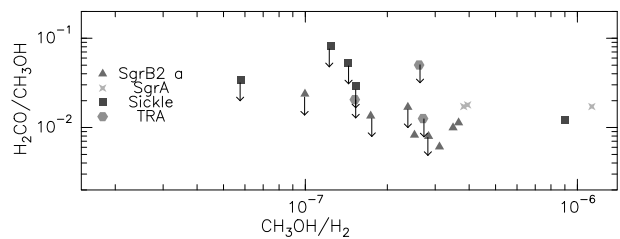}
\caption{Same as Fig. \ref{abun1} but for H$_2$CO.}\label{h2co}
\end{figure}

Because CH$_3$OH is the only molecule that has been detected in all sources and
it is believed to play a central role in the formation of some of the
observed complex organic molecules, we compared the abundances of the rest
of the complex molecules with that of CH$_3$OH.
Figure \ref{abun1} summarizes the results illustrating the abundance of
all the observed molecules relative to that of CH$_3$OH as a function of the 
CH$_3$OH abundance. As already mentioned, we have only included the significant 
upper limits in this figure.
The main results are:\\ 

--In spite of the CH$_3$OH's
abundance changing by a factor of $\sim$50, all the complex organic molecules, 
except that of the CH$_3$COOH with only upper limits to its abundance, show a 
surprisingly constant relative abundance with respect to CH$_3$OH.\\ 

--The mean [C$_2$H$_5$OH/CH$_3$OH] abundance ratio of $\sim$3.6$\times$10$^{-2}$ 
is constant within a factor 3.7.
The Sickle and the TRA clouds show upper limits to the [C$_2$H$_5$OH/CH$_3$OH]
ratios, which are a factor of $\ga$6 times lower than the mean ratio found in 
the typical GC clouds, a factor of $\ga$1.6 higher than the scattering of the 
data.\\

--The abundance ratios [(CH$_3$)$_2$O/CH$_3$OH] and [HCOOCH$_3$/CH$_3$OH]
present similar mean values of $\sim$3.9$\times$10$^{-2}$. These ratios show a
somewhat larger dispersion than that of C$_2$H$_5$OH but are constant within a 
factor of 7. \\

--For the [HCOOH/CH$_3$OH] ratio, we obtained a mean value of
$\sim$7.9$\times$10$^{-3}$ and a dispersion of a factor of 5.3.\\

--CH$_3$COOH was not been detected in the GC clouds but its 
\mbox{[CH$_3$COOH/CH$_3$OH]} ratio must be lower than $10^{-3}$. \\

--We also show in Fig. \ref{h2co} the [H$_2$CO/CH$_3$OH] ratio as a function
of the CH$_3$OH abundance. As for the more complex organic molecules, we find 
a constant ratio of $\sim$1.1$\times$10$^{-2}$, within a factor 3.  
\begin{figure}[h!]
\includegraphics[angle=270,width=8cm]{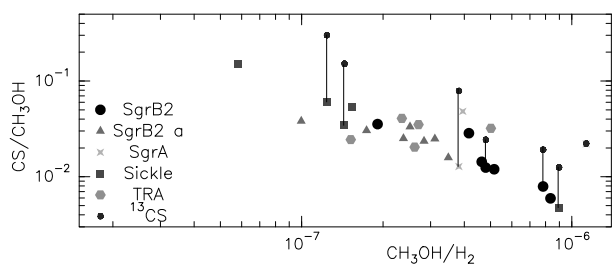}
\caption{Variation of [CS/CH$_3$OH] as a function of CH$_3$OH abundances.
The vertical bars show the changes due to the opacity.}\label{css}
\end{figure}

The [CS/CH$_3$OH] abundance ratio is shown in Fig. \ref{css}. This abundance
ratio decreases by a factor 60 as the CH$_3$OH abundance increases. This is the
only molecule for which we find a systematic trend, and the variation is clearly 
larger than those found for the complex organic molecules. As previously
mentioned, our CS column densities could be affected by opacity effects. To 
show the effects of opacity, we have also included the abundance ratio obtained
using the rarer CS isotopomer $^{13}$CS in Fig. \ref{css} as vertical bars.
The [CS/CH$_3$OH] abundance ratio derived from the optically thin 
isotopomer increases by a relatively constant value for all clouds independently
of  the CH$_3$OH abundance. 
The variation observed in the [CS/CH$_3$OH] ratio as a function of the CH$_3$OH
abundance, using the $^{13}$CS data, is 
similar or even larger than that derived from the CS data. We therefore consider
that the difference between the behavior of CS and the complex organic molecules
is real and related to the different type of chemistry in the GC clouds. 

\section{Comparison with Galactic disk objects}
\label{oobj}
In dark clouds, where the kinetic temperature 
is only 10$\,$K, the abundance of complex organic molecules is very low. 
In the dark cloud TMC-1, the CH$_3$OH, H$_2$CO, and HCOOH abundances are as low
as 2$\times$10$^{-9}$, 2$\times$10$^{-8}$, and 3$\times$10$^{-10}$, respectively
\citep[see e.g.][]{tur00}. For other dark clouds like L183, one finds similar 
abundances of  
8$\times$10$^{-9}$ for CH$_3$OH, of 2$\times$10$^{-8}$ for H$_2$CO, and
of 3$\times$10$^{-10}$ for HCOOH \citep{tur99,dic00}. 
Other complex
molecules, like C$_2$H$_5$OH, (CH$_3$)$_2$O, CH$_3$COOH, and HCOOCH$_3$, have 
not been detected  so far, but model predictions suggest C$_2$H$_5$OH abundances
of 10$^{-12}$ \citep{has93}.
High abundance of the complex organic molecules were first observed in the 
short-lived ($10^{5-6}\,$years) objects associated with 
massive star-forming regions known as hot cores.
There, it is believed that grain mantle evaporation release complex 
molecules, like alcohols, into gas phase. 
Recently, \citet{bot04a,bot04b,bot06} and \citet{kua04} detected 
relatively large
abundances of complex organic molecules, like HCOOCH$_3$ and
(CH$_3$)$_2$O, toward the so-called $``$hot corinos$"$ associated with warm
condensation surrounding low mass proto-stars.

In the following sections, we compare the abundances of complex organic molecules
in the GC with those found in the hot cores and hot corinos associated with star
formation.   

\subsection{Hot cores}
\label{oobj1}
For the comparison between the GC and hot cores associated to massive star
formation, we
selected the homogeneous data set by \citet{ike01} for several hot
cores (W51, NGC6334f, G327.3$-$0.6, G31.41+0.31, G34.3+0.2, G10.47+0.03, and the
Orion Hot Core). We also used data from \citet{rem02,rem03} to 
obtain the CH$_3$COOH abundances derived from their [HCOOCH$_3$/CH$_3$COOH]
ratio. 
The hot cores abundances of complex molecules and the H$_2$ column densities are
given in Table \ref{hot}. Both the H$_2$ and molecular column densities 
were derived from observations with similar beam sizes. Then the derived
fractional abundances are independent of beam dilution. Since our comparation is
based on the relative abundance between complex molecules, our results are not
affected by the size of the hot cores. 
\begin{figure}[h!]
\includegraphics[width=8.50cm]{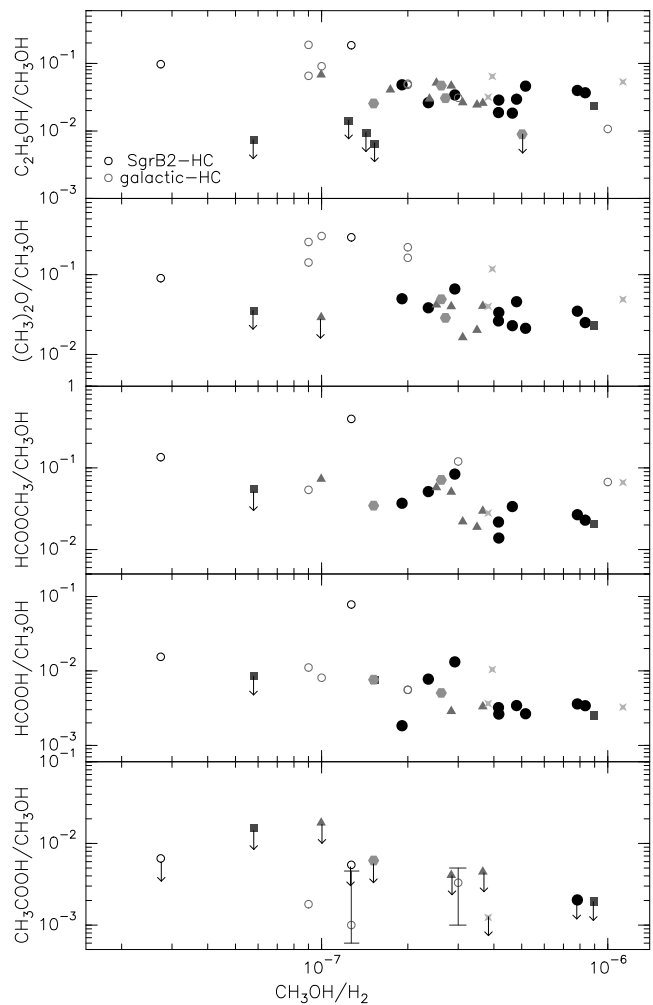}
\caption{Same as Fig. \ref{abun1} but including the hot core
abundances obtained from \citet{ike01}, \citet{rem02,rem03}, and
from our SGR~B2 data as open symbols.
\label{abun2}}
\end{figure}
We also used our data for 
the Sgr~B2N and Sgr~B2M hot cores in Table \ref{sou2}. Figure \ref{abun2} 
illustrates the comparison between the abundance ratios observed for the GC and 
those measured in hot cores.
As for the GC clouds, the abundance of CH$_3$OH in hot cores shows 
variations of more than one order of magnitude between different objects. 
Surprisingly, the abundance ratios relative to CH$_3$OH are also independent of 
the CH$_3$OH abundance and similar to those measured for the GC clouds.
The C$_2$H$_5$OH and HCOOH mean abundance ratios  with respect to CH$_3$OH 
between hot cores and GC clouds are different by less than a factor of 2.
The molecule that presents the largest difference in its abundance ratio with
respect to that in the GC clouds is (CH$_3$)$_2$O with a mean ratio of 
2$\times$10$^{-1}$, which is $\sim$6 times higher than those of the GC clouds. 
The [CH$_3$COOH/CH$_3$OH] ratios in hot cores agree with the upper-limit ratios
found in GC clouds.
Our data for the Sgr~B2N Hot Core shows slightly higher abundances of 
HCOOCH$_3$ and HCOOH than the rest of the hot cores. The Sgr~B2M Hot Core,
with the lowest CH$_3$OH abundance, shows similar ratios to those in the GC
sources.\\

  \begin{table*}
\begin{minipage}[t]{\textwidth}
\caption{Hot core abundances}             
\label{hot}      
\centering          
\begin{tabular}{l c c c c c c c c }     
\hline\hline       
source& $X_{\rm CH_3OH}$& $X_{\rm C_2H_5OH}$ 
& $X_{\rm (CH_3)_2O}$&
$X_{\rm HCOOCH_3}$  & $X_{\rm HCOOH}$& $X_{{\rm CH}_3{\rm COOH}}$& $X_{\rm H_2CO}$& N(H$_2$)  \\  
&$\times 10^{-7}$&$\times 10^{-8}$&$\times 10^{-8}$&$\times 10^{-8}$&$\times
10^{-9}$&$\times 10^{-9}$&$\times 10^{-8}$&$\times 10^{22}\rm cm^{-2}$\\
\hline       
High mass\footnote{\citet{ike01} and \citet{rem02,rem03}.}&&&&&&&&\\ 
W51~e1/e2       &3.0  	&0.9  	&        &3.0     &        &0.3--1.8    &	&36\\
NGC6334f        &2.0  	&0.9 	&4.0     &        &1.0     &		&	&20\\
G327.3$-$0.6    &1.0  	&1.0  	&3.0     &        &0.9     &        	&	&20\\
G31.41+0.31     &0.9  	&2.0  	&2.0     &        &1.0     &        	&	&16\\
G34.3+0.2       &0.9  	&0.6  	&1.0     &0.5     &        &0.2         & 	&30\\
G10.47+0.03     &2.0  	&1.0  	&3.0     &        &1.0     &        	&	&13\\ 
Omc HC          &10.0	&2.0  	&        &9.0     &        &        	&	&10\\ 
\hline
Low mass\footnote{\citet{bot06}.}&&&&&&&&\\
IRAS~4A		&$\la$0.1	&  	&$\la$2.8	&6.8		&4.6		&	&2.0	&160\\
IRAS~4B		&7.0		&  	&$\la$120.0	&220.0		&$\la$1000	&	&300.0	&8.1\\
IRAS~2A		&3.0		&  	&3.0		&$\la$134.0	&$\la$120	&	&20.0	&21\\
IRAS~16293	&1.0		&	&24.0		&34.0		&$\sim$62	&	&10.0	&7.5\\
\hline       
\end{tabular}
\end{minipage}
\end{table*}
It is remarkable that different objects like the hot cores and the GC clouds,
with different properties in terms of densities, H$_2$ column densities, and 
spatial extension, show such uniform abundance ratios of very 
complex molecules. Hot cores are denser than the GC clouds by more than 2 orders
of magnitude, and they typically have larger column densities than the GC
clouds. 
\subsection{Hot corinos}
\citet{bot04a,bot04b,bot06}, \citet{jor05} and \citet{kua04} observed the hot 
corinos associated with the low mass proto-stars, NGC1333~IRAS$\,$4A,
NGC1333~IRAS$\,$4B, NGC1333~IRAS$\,$2A and IRAS~16293-2422 
(A and B), which also show large abundances of HCOOCH$_3$, H$_2$CO, 
(CH$_3$)$_2$O and HCOOH. Table \ref{hot} summarizes the recent compilation 
of the hot corino abundances of complex organic molecules by \citet{bot06}.
In this case the molecular abundances were derived by assuming a source
size of 0.5$''$ for IRAS$\,$4A \citep{bot04a}, 0.25$''$ for IRAS$\,$4B, 0.43$''$
for IRAS$\,$2A \citep{mar04}, and 2$''$ for IRAS~16293-2422 \citep{caz03}.

Except for IRAS~4A, which shows lower abundances, HCOOCH$_3$, (CH$_3$)$_2$O, 
HCOOH, and H$_2$CO show higher abundances in hot corinos than in the GC 
molecular clouds and hot cores.
The data show that the abundance of CH$_3$OH varies between sources by nearly 
two orders of magnitude, 
similar to the variation observed in the GC clouds.  
As in the GC molecular clouds and in the hot cores, the abundance of the
complex organic molecules relative to that of CH$_3$OH is independent of the
CH$_3$OH abundance for different hot corinos \citep{bot06}.
 
Although the number of observations of hot corinos is still limited, it seems
that the abundance ratios, including species believed to be formed by
different chemistries, show a similar behavior to the GC sources. However,
the abundance of the molecules is very different, as expected if the grain
mantle composition in dark clouds were different than in the GC clouds and hot 
cores.

\section{Discussion}
With the exception of the clouds associated with the Sickle and the TRA, large 
abundances of organic complex molecules like CH$_3$OH, C$_2$H$_5$OH, 
(CH$_3$)$_2$O, HCOOCH$_3$, HCOOH, and H$_2$CO seems to be a general 
characteristic of molecular clouds in the GC at scales of a few hundred 
parsecs.  
The main characteristic of the GC molecular gas is that the complex organic
molecules show a constant relative abundance with respect to CH$_3$OH, within a
factor 4--8 over the whole range of the CH$_3$OH abundances in the observed GC
clouds. 
This contrasts with the [CS/CH$_3$OH] abundance ratio that  
increases by a factor of 60 when the CH$_3$OH abundance decreases.\\ 

We now discuss the origin of the chemistry of the complex
molecules found in the GC sources, including the low
[C$_2$H$_5$OH/CH$_3$OH] abundance ratio measured in the Sickle and the TRA
clouds. 
\begin{table*}
\begin{minipage}[t!]{\textwidth}
\caption{Summary of chemical models and observations of complex organic
molecules}             
\label{mod}      
\centering          
\begin{tabular}{l c c c c c c c }     
\hline\hline       
source& T& n(H$_2$)&$\frac{\rm C_2H_5OH}{\rm CH_3OH}$\footnote{The relative abundances of C$_2$H$_5$OH, 
(CH$_3$)$_2$O, HCOOCH$_3$, and H$_2$CO with respect to CH$_3$OH predicted by 
time-dependent chemical models, refer to the age of maximum abundance of the 
CH$_3$OH's daughters.}& 
$\frac{\rm (CH_3)_2O}{\rm CH_3OH}$$^a$&$\frac{\rm HCOOCH_3}{\rm CH_3OH}$$^a$&
 $\frac{\rm H_2CO}{\rm CH_3OH}$$^a$&age\\
& (K)& (cm$^{-3}$)&$\times10^{-2}$& $\times 10^{-2}$&$\times 10^{-2}$&
 &$\times10^5$y\\  
\hline       
Galactic center clouds         		&$\sim$100\footnote{From \citet{hut93a} and \citet{nem01a}. }&$\sim$10$^{4}$-10$^{5}$ &$\sim$3.66&$\sim$3.87&$\sim$3.96	&$\sim$1.10$\times$10$^{-2}$&       \\
hot cores\footnote{Mean abundance ratios for the hot core sources used in this work.}		&$\sim$100	&$\sim$10$^{6}$ &$\sim$6.90&$\sim$21.80&$\sim$8.03	&       &       \\
\citet{leu84}\footnote{Model 1 of the referred paper.}	&10     &10$^{4}$               &0.14  				&6.00 			&              		& 43.6			& $3.16$\\
\citet{her86}\footnote{Low-metal model in the referred paper.}	&10     &10$^{4}$               &2.18  				&5.25 			&              		&9.75			& $3.20$\\
\citet{her89}$^e$	&10     &$10^{4}$               &0.99  				&0.12 			&              		&65.5			& $1.00$\\
\citet{mil91}$^d$	&70     &2$\times 10^{5}$       &0.19  				&6.32$\times$10$^{-2}$  &1.26  			&1.18$\times$10$^{3}$ 	& $1.60$\\
\citet{has93}\footnote{Model starting with all the H in H$_2$.}&10     &$10^{4}$               &0.77  				&1.56$\times$10$^{-2}$  &       		&4.06$\times$10$^{2}$	& $3.20$\\
\citet{case93}-CR\footnote{Model fixed with the Orion Compact ridge observations.}	&100    &10$^{6}$        	&1.5$\times$10$^{-4}$  		&0.21 			&4.67$\times$10$^{-3}$  &1.05			&$1.00$\\ 
\citet{char95}           	&100    &2$\times$10$^{6}$       &       			&$\sim$2		&$\sim$1.5		&4$\times$10$^{-2}$	&$\sim$0.30\\ 
\citet{rod01}\footnote{Model without NH$_3$.}&100    &$10^{7}$               &3.75   			&30.00   		&10.00   		&        		&$0.12$\\ 
\citet{rod01}\footnote{Model with NH$_3$.}&100    &$10^{7}$               &20.00  			&12.50  			&0.50   		&        		&$0.60$\\ 
\citet{hor04}\footnote{Model 3 of the referred paper.}	&100    &$10^{6}$               &       		&0.75   		&3.33$\times$10$^{-2}$  &1.56	& $0.80$\\
\citet{pee06}\footnote{Model with (CH$_3$)$_2$O created only in gas phase.}	&100    &$10^{7}$               &       			&50.00  			&         		&       		& $0.20$\\
\hline       
\end{tabular}
\end{minipage}
\end{table*}

\subsection{Models for gas-phase formations of organic molecules}
Table \ref{mod} shows a compilation of the models proposed to explain the
formation of complex organic molecules. Since gas-phase chemistry cannot account
for the large abundance of alcohols, such as CH$_3$OH and C$_2$H$_5$OH, or for
other organic molecules like the H$_2$CO measured in hot cores, evaporation from
grain mantles has been proposed
as the main formation mechanism. Most of these models assume that  
(CH$_3$)$_2$O and HCOOCH$_3$ were produced in gas phase after the ejection of
CH$_3$OH. 
However, it has been recently claimed that HCOOCH$_3$ cannot be produced
efficiently in gas phase from CH$_3$OH \citep{hor04}. The large abundances
of this molecule found in the GC clouds clearly indicate that 
HCOOCH$_3$ is hardly likely to be produced in gas phase. However, the most
recent model from \citet{pee06} studied the possibilities for the formation of
(CH$_3$)$_2$O on grain mantles and concluded that the gas-phase reactions
are the dominant way to form (CH$_3$)$_2$O, while grain mantle reactions are
the minor source. 

From the theoretical point of view, so far, only (CH$_3$)$_2$O seems to be
produced in gas phase from reactions involving CH$_3$OH on relatively short 
time scales. 
Assuming that the gas phase models for the (CH$_3$)$_2$O formation are correct 
(see references in Table \ref{mod}), one would expect large changes in the
relative abundances between the daughter ((CH$_3$)$_2$O) and parent (CH$_3$OH) 
molecules as a function of time after ejection/evaporation. One expects low
[(CH$_3$)$_2$O/CH$_3$OH] abundance ratio in early times just when large 
abundances of CH$_3$OH are ejected into the gas phase and the gas phase 
processing starts. Low [(CH$_3$)$_2$O/CH$_3$OH] ratios are also expected in late
times in the low CH$_3$OH abundance regime when an important fraction of
CH$_3$OH has been converted into daughter molecules and both molecules have been
destroyed by gas-phase reactions. 
Between these two regimes, one would expect the [(CH$_3$)$_2$O/CH$_3$OH] ratio to
reach its maximum values, because the daughter molecules reach their maximum
abundance. \\

From the hot core models, one expects variations in
[(CH$_3$)$_2$O/CH$_3$OH] abundance ratio up to three orders of magnitude as a
function of time. 
This contrasts with our data for the GC clouds and the hot cores, which show
variations of less than one order of magnitude. Assuming that the chemical
models for hot cores are applicable in the GC, the only possibility of
explaining the high abundances of complex molecules and
the rather constant [(CH$_3$)$_2$O/CH$_3$OH] abundance ratios is to consider 
that all GC clouds in our study, distributed over 200$\,$pc, have undergone the
ejection of CH$_3$OH from grain mantles nearly simultaneously, 
$\sim$10$^{5}$ years ago. This possibility seems very unlikely, suggesting that
(CH$_3$)$_2$O is also ejected from grain mantels. 
Since (CH$_3$)$_2$O shows similar abundances with respect to CH$_3$OH than 
the other molecules that can only be formed on grain mantles like C$_2$H$_5$OH
and HCOOCH$_3$, our data support the scenario in which all complex organic 
molecules in the GC clouds, the hot cores, and also in hot corinos have been 
ejected/evaporated from grains. \\

Previously we also found differences in the behavior of the relative abundances 
of CS and the complex organic molecules.
This behavior is supported by the data in shock regions associated with 
outflows, as in L1448 \citep{jim05} and L1157 
\citep{bac97}. The CS abundance is
enhanced by the shock, but the enhancement of CH$_3$OH is $\sim$1 order of 
magnitude larger than that of CS.
This effect of differential ejection of the molecules can explain the variation 
in the complex organic molecules abundances in gas phase, while the CS abundance
remains marginally affected.

\subsection{Grain mantle erosion by shocks and time scales for depletion}
In the GC clouds, the typical dust-grain temperatures are low
\citep[T$\sim$10--30$\,$K,][]{nem04} and the ejection of 
molecules to gas phase cannot be produced by evaporation like in hot cores, but
by shocks with moderate velocities. 
The shock heats the gas and produces the sputtering of
molecules on the grain mantles. 
For shock velocities of $\sim$20$\,$km s$^{-1}$, the gas
temperature can reach $\sim$1000$\,$K \citep{kau96}. After the
shock passage, the gas rapidly cools on time scales of
$\sim$10$^{4}$ years for densities of 10$^5\,$cm$^{-3}$. 
The complex molecules ejected from the grains will again stick to the grains or
grain mantles. The time scales will depend  on the sticking coefficient.
A sticking coefficient of $\sim$1 is expected for mean
velocities corresponding to the thermal velocity at temperatures of
$\lesssim$200$\,$K because the complex molecules will be absorbed after every 
collision with a grain \citep{tie82}.
However, the sticking coefficient decreases when increasing the relative
velocity between the grains and the molecules \citep{lei85}.  \citet{buc91} and 
\citet{mas98} estimated that the sticking coefficient of
H is a function of the mean velocity.

The key parameter is the adsorption energy of the complex
molecules, which will depend on the surface of the dust grains. 
Since the CH$_3$OH 
abundance in gas phase is relatively large, grain mantles are completely eroded
and the most likely grain surfaces in the GC clouds would be graphite and 
silicates. The absorption energy of CH$_3$OH on carbon basal planes is 1600~K 
\citep{aik96} and on silicates is 2065~K \citep{all77}.  
For the typical mean velocities in the GC of 2--4$\,$km s$^{-1}$ (see next
section) the sticking coefficient will decrease and the time scale for sticking
would be $\sim$10$^5\,$years on silicates and graphites.
Similar time scales are found for the depletion of
C$_2$H$_5$OH, HCOOCH$_3$, and (CH$_3$)$_2$O molecules with absorption energies
$\sim$3000~K on silicates \citep{all77}.
Short time scales for depletion, like those expected in the GC, would require a 
mechanism that continuously ejects molecules from the grain mantles.

\subsection{Origin of the shocks in the GC}
\citet{nem01a,nem04} have summarized the possible 
mechanisms for the heating and the chemistry of the molecular gas in the GC. 
Turbulence and low-velocity shocks are proposed as the most likely mechanism for
driving the chemistry.
One of the key parameters for constraining the origin of the shocks is the 
velocity dispersion in the GC molecular clouds. 
High angular-resolution observation of the envelope of Sgr~B2 indicates that 
the warm molecular gas is highly turbulent with linewidths of $\sim$4$\,$km 
s$^{-1}$ \citep{marpin99}. \citet{smi00} show 
that MHD turbulence creates a wide range of shock velocities, but the larger 
amount of them should be produced with Mach numbers between 2--4. In the GC 
regions the sound speed is $\sim$1$\,$km~s$^{-1}$ (at 100$\,$K), and then the 
more abundant shocks associated with turbulence should have velocities between 2
and 4$\,$km s$^{-1}$, in agreement with the linewidth measured from high
angular-resolution observations. As discussed in the previous section, for a 
mean velocity of $\sim$3$\,$km s$^{-1}$, the depletion will be $\sim$10$^5$
years.

The energy of any kind of turbulence will decay if there is not any source 
injecting new energy. It is interesting to compare the depletion time scales
with the turbulence decay time scales to establish if the turbulence can
maintain mean velocities that are high enough to match the depletion time 
scales.  
The turbulence decays as $t^{-\alpha}$, where 
$\alpha$$\sim$0.8--1.0 \citep{mac98,sto98}. The decay time scale of these 
turbulences is similar to one dynamical time 
\citep[$t_{dyn}$=$L/\sigma_v$, where L is the turbulent 
length and $\sigma_v$ the turbulent velocity,][]{avi01,elm04}, for both HD and 
MHD turbulence. 
The dynamical time for turbulence in the GC, assuming a turbulent length of
5$\,$pc and a turbulent velocity dispersion of $15\,$km s$^{-1}$ (observed GC
linewidths), is $\sim$3$\times$10$^{5}\,$years, close to the depletion time
scale.  
Since we do not observe large depletions in our data, the existence of faster 
shocks with velocities high enough to produce the sputtering of the grain
mantles and with the time scales between shocks similar to the time scales for
depletion, $\sim$10$^5\,$years is required. Shocks with velocities of
$\ga$6$\,$km s$^{-1}$ are required in order to sputter molecules from the icy 
mantles on graphite and silicate grains by heavy atoms 
\citep{dra79}. Then, if due to a sudden event, 
the turbulence generated by this event could explain the presence of complex 
organic molecules in the GC clouds for $\sim$10$^5$ years.

It is so far unclear what produces the supersonic turbulence and/or frequent 
shocks in the GC. The high-velocity shocks and the presence of a highly
turbulent medium in the GC could be due to: a) the kinematics of the gas
subjected to the barred potential of our Galaxy, b) the wind-blown bubbles 
produced in evolved massive stars like supernovas, or c) cloud-cloud 
collisions. 
The time scales needed in the GC to maintain 
the large abundances are much smaller than those associated to the 
quasi-circular orbits ($\sim$1$\times$10$^{7}$~years), and a large number of
large scale shocks would be required to explain the observed properties of the 
complex molecules. The other possible
mechanisms could explain the observations if cloud-cloud collision and energetic
events driven by massive stars occur on time scales of 10$^5$ years. In
the scenario of a recent star burst in the GC as proposed to explain the fine 
structure lines of ionized gas \citep{nem05}, 
frequent energetic events associated to massive stars are 
expected to produce shocks with moderate velocities of $\sim$10$\,$km s$^{-1}$
\citep{marpin99} ejecting complex molecules to gas phase and a 
large amount of energy into the ISM.  
 
\subsection{The chemistry of complex molecules in the Galaxy}
The proposed gas-phase chemistry for some complex 
molecules does not seem to explain the similar abundances ratios observed in the
GC clouds, in hot cores, and in hot corinos. From the empirical data
gathered there, we can conclude that the most
likely explanation for the large abundance of  CH$_3$OH, HCOOH, HCOOCH$_3$,
(CH$_3$)$_2$O and C$_2$H$_5$OH is that all these molecules are ejected from
grains. It is interesting to note that in this scenario, the ice-mantle
composition in complex organic molecules in the GC and in hot cores must be very
similar. 
This is surprising in view of the changes of the complex organic molecules
abundances observed in the hot corinos. The presence of complex molecules on
grains can be due to grain chemistry or to the depletion after formation in gas
phase.

Further investigations of the abundance of organic
molecules in dark clouds are needed to constrain the molecules formed on ice 
mantles by grain chemistry in different types of molecular clouds.

\subsection{Sources with low C$_2$H$_5$OH abundances}
We also observed some clouds where the relative abundance of some organic 
complex molecules like C$_2$H$_5$OH seems to be significantly smaller than 
expected from these observed in other sources in the GC. 
These sources are the Sickle (\mbox{MC G+0.18+0.04} and \mbox{MC G+0.20+0.03})
and the -30$\,$km s$^{-1}$ velocity component of the TRA (\mbox{MC
G+0.13+0.02}). We detected high CH$_3$OH abundances of $\sim$10$^{-7}$, but only
significant upper limits, $\la$0.4-5$\times$10$^{-9}$ to the C$_2$H$_5$OH
abundance.  
The stars from the Quintuplet and the Arches Clusters are said to heat and 
ionize the Sickle and the TRA \citep{nem01b} creating
photodissociation-regions (PDRs) around them.
In particular, the \mbox{-30$\,$km s$^{-1}$} component is clearly associated
with the TRA \citep{segu87}.\\

The chemistry in these sources is then expected to be strongly affected by the
presence of the UV radiation. It is very likely that the complex molecules in 
these sources are ejected to gas phase by shocks with the same abundances as 
observed in other GC molecular cloud. Additional evaporation from grain mantles 
can be ruled out since the heating will require UV photons that will also
destroy the complex molecules. Due to a large UV radiation-field, the 
relative abundance of the complex molecules will be affected by 
 dissociation. 
One can make a rough estimate of the effects of the UV radiation on the 
molecular abundances by comparing their photodissociation rates. 
Using the UMIST RATE99 \citep{let00} and the Ohio State University 
databases\footnote{www.physics.ohio-state.edu/$\sim$eric/research.html}, we 
found that the photodissociation rates of (CH$_3$)$_2$O and 
C$_2$H$_5$OH are larger than those of CH$_3$OH, HCOOH and H$_2$CO.
\begin{figure}[h!]
\includegraphics[width=8cm]{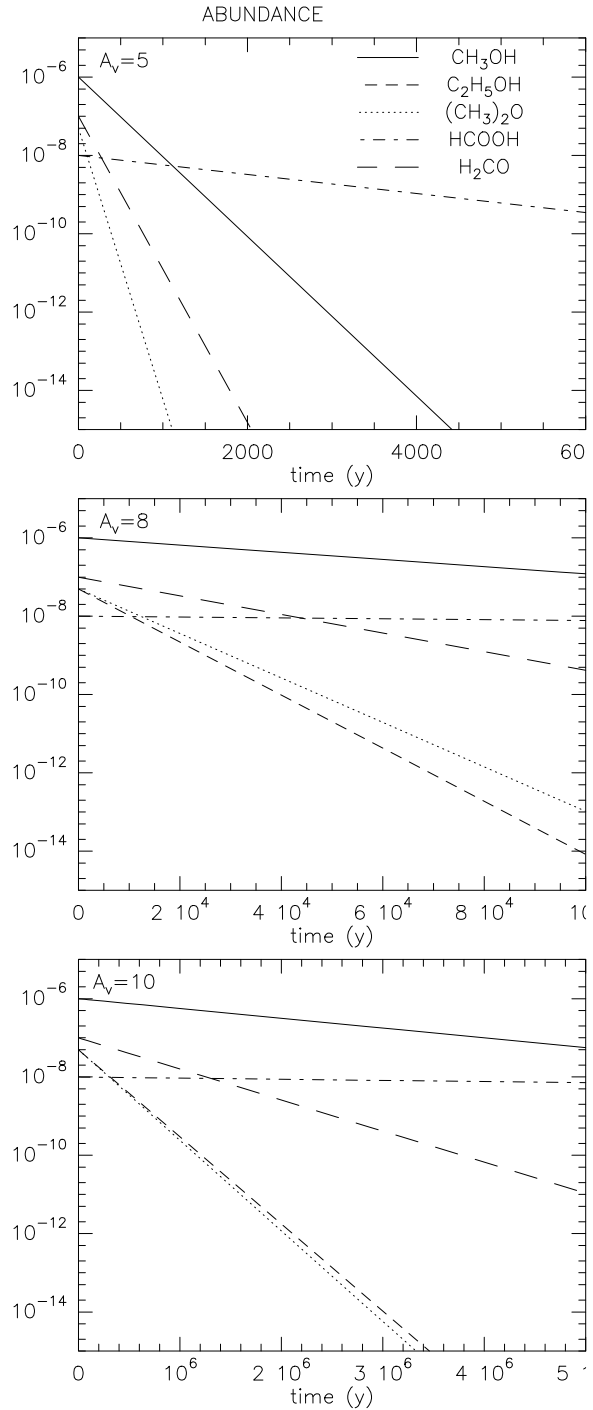}
\caption{Photodestruction of different complex molecules as a function of time 
for different visual extinctions (A$_{\rm v}$). The initial molecular abundances for the
molecules are considered
to be those typically measured in the observed GC clouds.
G$_0$=10$^3$, observed by \citet{nem04}, is used for the 
UV radiation field.\label{fot1}}
\end{figure}

In Fig. \ref{fot1} we show the results of a simple model in which we
represent how the ejected complex organic molecules are photodestroyed as a 
function of time and visual extinction. The model considers a far-ultraviolet 
incident field 10$^3$ times higher than in the local ISM (G$_0$=10$^3$), which
is the typical radiation field observed in GC PDRs \citep{nem04}. 
For the regions with low 
$A_{\rm v}$, all the abundances decrease, but the abundance of C$_2$H$_5$OH
decreases faster than the other organic complex molecules. Obviously the time
scale for photodissociation depends on the visual extinction. We estimate a
visual extinction of $A_{\rm v}$$\sim$4--17$\,$mag for the Sickle and the TRA
using our H$_2$ column densities in Table~\ref{sou2} and the conversion factor
given by \citet{boh75}, $A_{\rm v}$(mag)=10$^{-21}\times$N$_{H_2}$ (cm$^{-2}$).
The photodissociation 
times scales for CH$_3$OH will be comparable to depletion time scales for an
averaged visual extinction of $\sim$8$\,$mag. 
\begin{figure}[h!s]
\includegraphics[width=8cm]{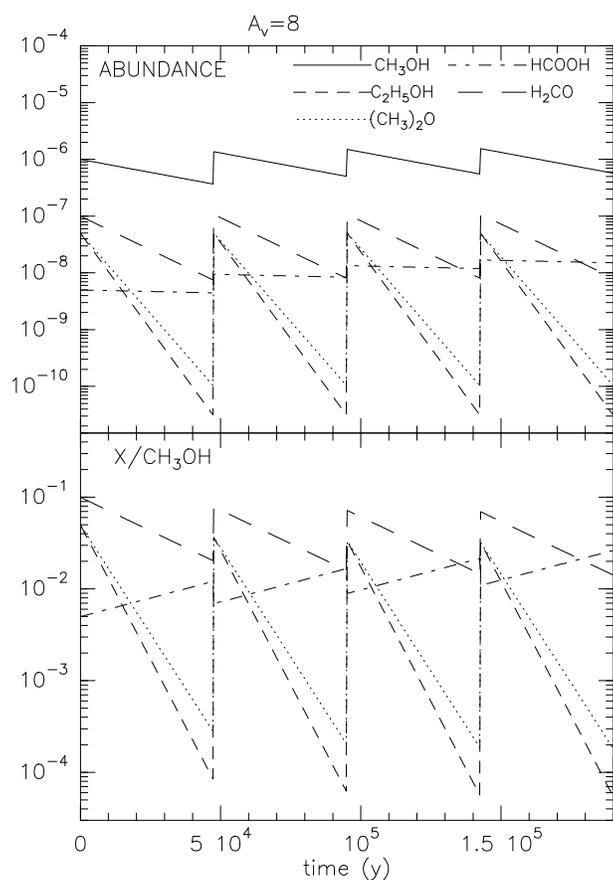}
\caption{Time evolution of the abundance (top) and relative abundance (bottom) of complex
organic molecules with respect to that of CH$_3$OH in a PDR with
$A_{\rm v}$=8~mag where a periodic shock every $5\times10^{4}\,$years  
enhances the abundances to their initial
values. The radiation field is G$_0$=10$^3$.\label{fot2}}
\end{figure}

For this visual extinction the other complex 
organic molecules will photodissociate much faster than the depletion time
scale.  
This suggests that in the presence of UV radiation, large abundances of
molecules like C$_2$H$_5$OH would require more frequent shocks than in the
shielded regions. This is illustrated in Fig. \ref{fot2}, where we present the
results of a simple model to show the time evolution of complex organic
molecules in a region of $A_{\rm v}$=8$\,$mag.
In this model, we consider the effects of the UV radiation on the complex
molecules that have been ejected by periodic shocks of 
5$\times10^{4}\,$years. 
They are ejected with the abundances
measured in the GC and are destroyed because of the photodissociation. 
This simple model reproduces the constant high abundances observed 
for CH$_3$OH, while the abundances of C$_2$H$_5$OH can vary by $\sim$3$\,$~orders
of magnitude. In agreement with the observations, the photodissociation can 
reproduce the low abundances of C$_2$H$_5$OH in specific regions in the 
GC clouds. Further observation of complex organic molecules toward PDRs in the
Galactic disk and in the Galactic center are required to test the proposed 
scenario. 

\section{Conclusion}
We have found very high abundances for the complex organic molecules 
CH$_3$OH, C$_2$H$_5$OH, HCOOCH$_3$, (CH$_3$)$_2$O, and HCOOH in the GC
regions, similar to or even higher than those measured in Galactic hot cores.
The abundance ratios of these molecules are fairly constant in all the GC
clouds.  
The most likely explanation for these large abundances is that they have been 
ejected from grain mantles by shocks.
The highly turbulent molecular clouds in the GC could help to maintain the
large gas phase abundance of complex molecules for time scales of $\sim$10$^5$
years after the ejection by shocks.  
The rather uniform abundance ratios in clouds distributed over 200 pc 
indicate that the average composition of grain mantles is similar for all 
the GC clouds and also for the hot cores in the Galactic disk.
We found that, in the GC PDRs, UV photons can differentially photodissociate the
complex molecules, changing their relative abundances dramatically. 

\begin{acknowledgements}
We wish to thank S. Bottinelli and her co-authors for communicating the results
of their work prior to publication. This work was supported by the Spanish
Ministerio de Educaci\'on y Ciencia under projects AYA 2002-10113-E,
AYA~2003-02785-E, and ESP~2004-00665. 
\end{acknowledgements}

\clearpage
{\headsep 6cm
\begin{landscape}
\begin{longtable}{l c l c c c c c c c c c }     
\caption{Relative abundances and H$_2$ column densities\label{sou2}}\\      
\hline
\hline
source&$V_{\rm rad}$&$T_{\rm rot}$&$X_{\rm CH_3OH}$&$X_{\rm C_2H_5OH}$&$X_{\rm (CH_3)_2O}$&
$X_{\rm HCOOCH_3}$&$X_{\rm HCOOH}$&$X_{\rm CH_3COOH}$&$X_{\rm H_2CO}$&$X_{\rm CS}$ & N(H$_2$)\\  
&(km s$^{-1}$)&(K)&$\times10^{-7}$&$\times10^{-8}$&$\times10^{-8}$&
$\times10^{-8}$&$\times10^{-9}$&$\times10^{-9}$&$\times10^{-9}$&$\times10^{-9}$&
$\times10^{22}$cm$^{-2}$\\ 
\hline
\hline
\endhead
\endfoot
MC G$-$0.96+0.13  & 140    &8.0\footnote[1]{Derived from CH$_3$OH.}  				&10.6 &  $\la$ 1.4    & $\la$ 8.3 & $\la$10.7       & $\la$12.2 &  $\la$30.0 &     & 16.6 &  0.2\\
MC G$-$0.55$-$0.05  & $-$93    &8.0$^1$  				& 1.6 &  $\la$ 1.8    & $\la$ 0.6 & $\la$ 2.7       & $\la$ 1.0 &  $\la$ 8.5 &     &  3.6 &  1.5\\
MC G$-$0.50$-$0.03  & $-$92    &8.0$^1$  				& 0.4 &  $\la$ 1.0    & $\la$ 0.3 & $\la$ 1.1       & $\la$ 0.4 &  $\la$ 3.6 &     &  1.6 &  2.5\\
MC G$-$0.42$-$0.01  & $-$71    &8.0$^1$  				& 0.6 &  $\la$ 1.9    & $\la$ 0.6 & $\la$ 2.9       & $\la$ 1.1 &  $\la$ 9.5 &     &  3.5 &  0.6\\
MC G$-$0.32$-$0.19  & $-$70.4  &8.0$^1$  				& 0.9 &  $\la$ 0.8    & $\la$18.1 & $\la$ 6.2       & $\la$23.4 &  $\la$17.5 &     & 10.0 &  0.2\\
MC G$-$0.32$-$0.19  & $-$24    &8.0$^1$				& 0.2 &  $\la$ 0.6    & $\la$ 6.3 & $\la$ 4.5       & $\la$17.0 &  $\la$12.7 &     &  7.3 &  0.2\\
MC G$-$0.32$-$0.19  & 26     &8.0$^1$  				& 1.1 &  $\la$ 2.6    & $\la$27.0 & $\la$ 19.5      & $\la$73.3 &  $\la$54.7 &     & 24.5 &  0.07\\
MC G$-$0.11$-$0.08  & 19.8   &12.0\footnote[2]{Derived from C$_2$H$_5$OH.}-10.0$^1$    	&11.3 &         6.0   &    5.6 &  7.5       	  &       3.7 &  $\la$ 3.9 &19.5        & 25.0/43.4 \footnote[3]{From $^{13}$CS  \citep{mar06}.}&  1.0\\
MC G$-$0.08$-$0.06  & 29.6   &10.0$^2$-12.0$^1$     & 4.0 &         2.5   &       4.7 & $\la$ 2.6       &       4.1 &  $\la$ 4.5 &7.1         & 19.0 &  0.6\\
MC G$-$0.02$-$0.07  & 47     &10.8\footnote[4]{Derived from HCOOCH$_3$.}-14.0$^2$	& 3.8 &1.2&       1.5 &     1.1         &       1.4 &  $\la$ 0.5 &6.6 	  & 4.9/17.5 $^3$&  6.8\\
MC G+0.04+0.03  & $-$26    &10.0$^1$   			& 2.7 &         0.8   &       0.8 & $\la$ 0.7       & $\la$ 0.6 &  $\la$ 2.0 &$\la$3.4    &  9.5 &  1.1\\
MC G+0.07$-$0.07  & 52     &10.0$^1$-12.8$^2$     & 2.6 &         1.2   &       1.3 &     1.9         &       1.3 &  $\la$ 1.9 &$\la$13.1   &  5.3 &  2.2\\
MC G+0.13+0.02  & $-$30    &8.0\footnote[5]{Assumed 8$\,$K.}  				& 5.0 & $\la$ 0.5     & $\la$ 1.4 & $\la$ 3.4       & $\la$ 2.1 &  $\la$ 9.4 &     & 16.1 &  0.6\\
MC G+0.13+0.02  & 52     &8.0$^5$  				& 2.4 &  $\la$ 0.5    & $\la$ 1.6 & $\la$ 3.7       & $\la$ 2.4 &  $\la$10.5 &     &  9.6 &  0.4\\
MC G+0.17+0.01  & 59     &10$^1$-12.7$^2$   	& 1.5 & 0.4           & $\la$ 0.3 &     0.5         &       1.2 &  $\la$ 0.9 &$\la$3.1    &  3.7 &  1.3\\
MC G+0.18$-$0.04  & 27.1   &8.0$^5$   				& 1.2 &$\la$ 0.2      & $\la$ 0.8 & $\la$ 0.9       & $\la$ 1.7 &  $\la$ 2.5 &$\la$10.2   &  7.5/37.3 $^3$ &  0.7\\
MC G+0.18$-$0.04  & 85.8   &8.0$^5$   				& 1.4 &$\la$ 0.1      & $\la$ 0.6 & $\la$ 0.7       & $\la$ 1.3 &  $\la$ 1.9 &$\la$7.8    &  5.0/22.0 $^3$ &  0.8\\
MC G+0.20$-$0.03  & 24.9   &8.0$^5$   				& 0.6 &$\la$ 0.04     & $\la$ 0.2 & $\la$ 0.3       & $\la$ 0.5 &  $\la$ 0.9 &$\la$2.0    &  8.7 &  1.7\\
MC G+0.20$-$0.03  & 87.7   &8.0$^5$   				& 1.5 &$\la$ 0.09     & $\la$ 0.4 & $\la$ 0.7       & $\la$ 1.1 &  $\la$ 2.0 &$\la$4.4    &  8.3 &  0.9\\
MC G+0.24+0.01  & 36     &12.6$^2$-8.0$^1$      & 9.0 &         2.1   &       2.1 &     1.8         &       2.3 &  $\la$ 1.7 &10.9        &  4.3/11.3 $^3$ &  3.0\\
MC G+0.62$-$0.10  & 56     &11.0$^2$-7.8$^1$     	& 3.5 &         0.8   &       0.7 &     0.7         & $\la$ 0.3 &  $\la$ 1.1 &3.5         &  5.5 &  2.6\\
MC G+0.64$-$0.08  & 63     &10.0$^2$$^1$  	&10.0 &         0.7   & $\la$ 0.3 &     0.7         & $\la$ 0.4 &  $\la$ 1.8 &$\la$2.4    &  3.8 &  2.0\\
MC G+0.67$-$0.06  & 51     &10.5$^2$  				& 3.1 &         0.8   &       0.5 &     0.7         & $\la$ 0.4 &  $\la$ 1.3 &1.9         &  7.8 &  3.1\\
MC G+0.68$-$0.10  & 21     &9.0$^2$-6.0$^4$     	& 2.5 &         1.3   &       1.1 &     1.5         & $\la$ 0.4 &  $\la$ 1.3 &2.1         &  8.4 &  3.2\\
MC G+0.70$-$0.01  & 62     &11.0$^2$-10.0$^1$    	& 3.7 &         0.9   &       1.5 &     1.1         &       1.2 &  $\la$ 1.6 &4.2         &&  2.8\\
MC G+0.70$-$0.09  & 43     &10.0$^1$     			& 1.7 &         0.7   & $\la$ 0.3 & $\la$ 1.3       & $\la$ 0.4 &  $\la$ 3.6 &$\la$2.4    &  5.3 &  2.9\\
MC G+0.71$-$0.13  & 40     &8.0$^1$  				& 2.4 & $\la$   0.8   & $\la$ 0.6 & $\la$ 2.8       & $\la$ 0.9 &  $\la$ 7.9 &$\la$0.6    &  6.0 &  2.8\\
MC G+0.76$-$0.05  & 27     &10.0$^1$$^4$      	& 2.8 &         1.3   &       1.1 &     1.5         &       0.8 &  $\la$ 1.2 &$\la$2.3    &  6.7 &  3.8\\
MC G+0.694$-$0.017& 66     &8.0$^1$-11.0$^4$     	& 8.3 &         3.1   &       2.1 &     1.9         &       2.8 &  $\la$ 2.4 &     &  5.0 &  3.9\\
MC G+0.693$-$0.027& 68     &10.0$^1$-11.5$^4$    	& 7.8 &         3.1   &       2.7 &     2.1         &       2.8 &  $\la$ 1.6 &     &  6.2/15.0 $^3$  &  4.1\\
MC G+0.627$-$0.067& 49     &10.0$^1$-6.0$^4$  	& 4.8 &         1.4   &       2.2 & $\la$ 0.8       &       1.6 &  $\la$ 2.2 &     &  6.0/11.7 $^3$  &  2.2\\
MC G+0.630$-$0.072& 46     &9.0$^1$-12.0\footnote[6]{Derived from (CH$_3$)$_2$O.}     	& 5.2 &         2.4   &       1.1 & $\la$ 0.9       &       1.4 &  $\la$ 2.5 &     &  6.2 &  1.4\\
MC G+0.672$-$0.014& 54     &10.0$^1$$^4$     	& 1.9 &         0.9   &       1.0 &     0.7         &       0.4 &  $\la$ 1.2 &     &  6.8 &  5.5\\
MC G+0.640$-$0.046& 58     &10.0$^1$-7.0$^6$     	& 4.2 &         1.2   &       1.4 &     0.6         &       1.1 &  $\la$ 1.6 &     & 11.9 &  4.7\\
MC G+0.635$-$0.069& 48     &11.0$^6$-8.0$^4$     	& 4.6 &         0.9   &       1.1 &     1.6         & $\la$ 0.6 &  $\la$ 2.5 &     &  6.7 &  2.3\\
MC G+0.659$-$0.035& 62     &11.0$^1$-16.0$^4$    	& 4.2 &         0.8   &       1.1 &     0.9         &       1.3 &  $\la$ 0.7 &     && 13.9\\
MC G+0.681$-$0.028& 65     &11.0$^1$-12.0$^4$   	& 2.9 &         1.0   &       1.9 &     2.5         &       3.9 &  $\la$ 0.8 &     &&  8.8\\
MC G+0.673$-$0.025& 68     &18.5$^4$-13.9$^1$   	& 2.4 &         0.6   &       0.9 &     1.2         &       1.8 &  $\la$ 0.8 &     && 14.6\\
SGR~B2N        & 65     &72.6$^2$ 				& 1.3 &         2.3   &       3.5 &     5.0         &       9.9 &  $\la$ 0.7 &     && 19.0\\
SGR~B2M        & 63     &55.4$^2$ 				& 0.3 &         0.3   &       0.3 &     0.4         &       0.4 &  $\la$ 0.2 &     && 65.0\\              
\hline                  
\end{longtable}
\end{landscape}}
\clearpage

\end{document}